\documentclass[fleqn,usenatbib]{mnras}

\usepackage{newtxtext,newtxmath}
\usepackage[T1]{fontenc}

\DeclareRobustCommand{\VAN}[3]{#2}
\let\VANthebibliography\thebibliography
\def\thebibliography{\DeclareRobustCommand{\VAN}[3]{##3}\VANthebibliography}

\usepackage{graphicx}	% Including figure files
\usepackage{amsmath}	% Advanced maths commands
\title[Periodicity of SGR J1935+2154 Bursts]{Revisit the periodicity of SGR J1935+2154 bursts with updated sample}

\author[Xie et al.]{Sheng-Lun Xie$^{1,2}$,
Ce Cai$^{2,9}$,
Shao-Lin Xiong$^{2}$\thanks{E-mail: xiongsl@ihep.ac.cn (SLX)},
Yun-Wei Yu$^{1}$\thanks{E-mail: yuyw@ccnu.edu.cn (YWY)},
Yan-Qiu Zhang$^{2,3}$,
Lin Lin$^{4}$,
Zhen Zhang$^{2}$,
\newauthor
Wang-Chen Xue$^{2,3}$,
Jia-Cong Liu$^{2,3}$,
Yi Zhao$^{2,4}$,
Shuo Xiao$^{2,3}$,
Chao Zheng$^{2,3}$,
Qi-Bin Yi$^{2,5}$,
Peng Zhang$^{2}$,
\newauthor
Ping Wang$^{2}$,
Rui Qiao$^{2}$,
Wen-Xi Peng$^{2}$,
Yue Huang$^{2}$,
Xiang Ma$^{2}$,
Xiao-Yun Zhao$^{2}$,
Xiao-Bo Li$^{2}$,
\newauthor
Shi-Jie Zheng$^{2}$,
Ming-Yu Ge$^{2}$,
Cheng-Kui Li$^{2}$,
Xin-Qiao Li$^{2}$,
Xiang-Yang Wen$^{2}$,
Fan Zhang$^{2}$,
Li-Ming Song$^{2}$,
\newauthor
Shuang-Nan Zhang$^{2}$,
Zhi-Wei Guo$^{2,6}$,
Xiao-Lu Zhang$^{2,7}$,
Guo-Ying Zhao$^{2,5}$,
Chao-Yang Li$^{2,8}$
\\
$^{1}$ Institute of Astrophysics, Central China Normal University, Wuhan 430079, China\\
$^{2}$ Key Laboratory of Particle Astrophysics, Institute of High Energy Physics, Chinese Academy of Sciences, 19B Yuquan Road, Beijing 100049, China\\
$^{3}$ University of Chinese Academy of Sciences, Chinese Academy of Sciences, Beijing 100049, China\\
$^{4}$ Department of Astronomy, Beijing Normal University, Beijing 100088, People’s Republic of China\\
$^{5}$ Department of Physics, Xiangtan University, Xiangtan, Hunan Province 411105, China\\
$^{6}$ College of physics Sciences Technology, Hebei University, No. 180 Wusi Dong Road, Lian Chi District, Baoding City, Hebei Province 071002, China\\
$^{7}$ College of Physics and Engineering, Qufu Normal University, Qufu 273165, China\\
$^{8}$ Physics and Space Science College, China West Normal University, Nanchong 637002, China\\
$^{9}$ College of Physics, Hebei Normal University, 20 South Erhuan Road, Shijiazhuang 050024, Hebei, China\\
}

\pubyear{2022}

\begin{document}
\label{firstpage}
\pagerange{\pageref{firstpage}--\pageref{lastpage}}
\maketitle

\begin{abstract}
Since FRB 200428 has been found to be associated with an X-ray burst from the Galactic magnetar SGR J1935+2154, it is interesting to explore whether the magnetar bursts also follow the similar active periodic behavior as some repeating FRBs. Previous studies show that there is possible period about 230 day in SGR J1935+2154 bursts. Here, we collected an updated burst sample from SGR J1935+2154, including all bursts reported by Fermi/GBM and GECAM till 2022 January. We also developed a targeted search pipeline to reveal more bursts from SGR J1935+2154 in the Fermi/GBM data from 2008 August to 2014 December (i.e. before the first burst detected by Swift/BAT). With this burst sample, we re-analyzed the possible periodicity of SGR J1935+2154 bursts using the Period Folding and Lomb-Scargle Periodogram methods. Our results show that the periodicity $\sim$238 day reported in literature is probably fake and the observation effects may introduce false periods (i.e. 55 day) according to simulation tests. We find that, for the current burst sample, the most probable period is 126.88$\pm$2.05 day, which could be interpreted as the precession of the magnetar. However, we note that the whole burst history is very complicated and difficult to be perfectly accommodated with any period reported thus far, therefore more monitoring observations of SGR J1935+2154 are required to test any periodicity hypothesis.
\end{abstract}


%%%%%%%%%%%%%%%%%%%%%%%%%%%%%%%%%%%%%%%%%%%%%%%%%%

%%%%%%%%%%%%%%%%% BODY OF PAPER %%%%%%%%%%%%%%%%%%

\section{Introduction}

Highly-magnetized neutron stars called magnetars are perceived as the nature of soft gamma-ray repeaters (SGRs) and anomalous X-ray pulsars (AXPs), which can intermittently produce gamma/X-ray bursts \citep{DuncanThompson1992,vanKerkwijk1995,Kouveliotou1998,Banas1997,KaspiBeloborodov2017}. On April 28th, 2020, it was discovered that a fast radio burst \citep[FRB 200428,][]{Li2021,Bochenek2020,CHIME2020b} was produced by the Galactic magnetar SGR J1935+2154 and this FRB was in temporal coincidence with a non-thermal X-ray burst from the magnetar \citep{Li2021,Mereghetti2020,Tavani2020,Ridnaia2021}.

FRBs are super bright radio pulses with a duration of a few milliseconds \citep{Lorimer2007,Thornton2013}, featuring in their abnormally high dispersion measures (DMs). 
Since the first discovery of the Lorimer burst from the pulsar data of Parkes radio telescope \citep{Lorimer2007}, over 600 FRBs have been found \citep{Petroff2016}, which enabled us to learn more characteristics of the FRBs, such as the localization of their host galaxies \citep{Chatterjee2017,Bannister2019,Ravi2019,Macquart2020,Marcote2020}, polarization, rotation measure and the highly magnetic environment around the sources \citep{Michilli2018}. According to their potential cosmological origin, FRBs have been widely connected to some violent activities and even catastrophic coalescences or collapses
of compact binaries or objects (see \cite{Platts2019}\footnote{\url{https://frbtheorycat.org}} for a review).

Although most FRBs were observed to be one-off burst, several repeating sources have been discovered, such as FRB 121102, FRB 180916, FRB 20190520B, FRB 20181030A, and FRB 20201124A \citep{Spitler2016,Chatterjee2017,Marcote2017,Tendulkar2017}. Undoubtedly, the repetition of these FRBs disfavors the models involving catastrophic events and, instead, strongly supports the activities of compact objects, in particular, young magnetars \citep{Popov2010,Kulkarni2014,Connor2016,Cordes2016,Katz2016,Lyutikov2017,Metzger2017,Cao2017}. This hypothesis has further been confirmed by the discovery of the association of FRB 200428 with an X-ray burst from SGR J1935+2154 \citep{Li2021,Mereghetti2020,Tavani2020,Ridnaia2021}. This discovery not only provides an important constraint on the physical mechanisms responsible for the FRB phenomena \cite[e.g.,][]{Lu2020,Margalit2020,Wu2020,Yang2020,Zhang2020b,Yu2021}, but also hints that we may use the X-ray bursts of magnetars to indirectly probe the temporal behavior of the repeating FRBs.

It is worthy to note that repeating FRBs could have periodic window behavior (PWB), e.g. FRB 180916 was found to have a possible period of 16.35 ± 0.15 days \citep{CHIME2020a}, and FRB 121102 could be with a period of 156 days \citep{Rajwade2020,Cruces2021}. Therefore, it is very interesting to investigate whether the activities of SGR J1935+2154, which has emitted FRB 200428, also have a PWB. This has been studied by some previous works: \cite{Denissenya2021D} argued a period of 231 days, using likelihood analysis with the data of IPN (Interplanetary Network) instruments from 2014 July to 2021 February; \cite{Grossan2021} estimated a possible period of about 231 days by analyzing the observations from IPN from 2014 July to 2020 May; \cite{Zou2021} suggested that the period is about 238 days using the data of Fermi/GBM (Fermi Gamma-ray Burst Monitor) from 2014 July to 2021 October.

In this paper, we carry out a comprehensive targeted search on X-ray bursts of SGR J1935+2154 using Fermi/GBM \citep{Meegan2009} data from 2008 August to 2014 December. We also collect the most recent burst samples from Fermi/GBM and all bursts detected by GECAM \citep[Gravitational wave high-energy Electromagnetic Counterpart All-sky Monitor,][]{LiWen2021,Xiao2022} during its first year of observation. With more bursts covering a longer time interval than previous studies \citep{Denissenya2021D, Grossan2021, Zou2021}, we hope to make a better constraint on the periodicity searching.
In Section \ref{sec:GTS}, we present the search process and report the burst candidates. In Section \ref{sec:pwb}, we use the Period Folding and Lomb-Scargle algorithm to search the possible periodic active window. Finally, the summary and discussion are given in Section \ref{sec:con}.

%%%%%%%%%%%%%%%%%%%%%%%%%%%%%%%%%%%%%%%%%%%%%%%%%%

\begin{table}
\caption{Spectral models used in the targeted search for X-ray bursts from SGR J1935+2154.}
\label{tab:template}
\begin{tabular}{lccc}
\hline
Spectrum Template & alpha & beta & Epeak or kT (keV) \\  \hline
Band              & 0.06  & -5.3 & 35    \\
BlackBody         & ...   & ...  & 9     \\
Comptonized       & 0.5   & ...  & 35    \\
OTTB              & ...   & ...  & 32    \\
Powerlaw          & -2    & ...  & 100   \\
\hline
\end{tabular}
\end{table}

\section{Targeted Search for Bursts}
\label{sec:GTS}

SGR J1935+2154 is a Galactic magnetar, which was first found by the Swift Burst Alert Telescope (BAT) in 2014 \citep{Stamatikos2014}. It has experienced four active episodes before 2020, respectively \citep{Younes2017,Lin2020a,Lin2020b}. Since 2021, there exist at least 3 active episodes so far.

Targeted sub-threshold search for bursts from SGR J1935+2154 is very important, because there are potential bursts which are rather weak and unable to trigger the instrument with normal threshold. Many works show evidence that some SGR bursts did not trigger detector \citep{Lin2020c,Lin2020a,Lin2020b,Mereghetti2020,Younes2020,Yang2021,Zou2021}. However, some of previous studies only searched SGR bursts after the first reported burst by Swift/BAT in 2014 and ignored the time interval before 2014.
In this paper, we implement a targeted search on Fermi/GBM Continuous high time resolution (CTIME) and Time-tagged events (TTE)\footnote{\url{https://heasarc.gsfc.nasa.gov/FTP/Fermi/data/gbm/daily/}} from 2008-08-12 to 2014-12-31 (UTC) with the coherent search method \citep{Blackburn2015,Cai2021}.

We have developed a pipeline to search for gamma-ray bursts (GRBs) and SGRs using the traditional signal-to-noise ratio (SNR) method for blind search and the coherent search method for targeted search \citep[e.g.,][]{Cai2021}. In this targeted search for SGR J1935+2154 bursts, we use 5 spectral templates (see Table \ref{tab:template}) and set the trigger threshold of Log-Likelihood Ratio (LR) to be 20. Thousands of burst candidates have been found, and we screened them according to the following criteria:

\begin{enumerate}

\item Excluding those candidate events when the spacecraft is near the South Atlantic Anomaly (SAA) or the location of SGR J1935+2154 is in the Earth shadow;
\item Excluding those events with the incident angle of most significant detectors greater than 60° and those events which are falsely triggered by the sharp variation of background count rates caused by instrument startup and shutdown. The number of these two kinds of events is about $\sim $7400;
\item Excluding the solar flares ($\sim $1800 events), particle events ($\sim $3800), Terrestrial Gamma Ray Flash \citep[TGF,][]{Briggs2013} and Terrestrial Electron Beam \citep[TEB,][]{Xiong2012}\footnote{\url{https://fermi.gsfc.nasa.gov/ssc/data/access/gbm/tgf/}}$^{,}$\footnote{\url{https://gammaray.nsstc.nasa.gov/gbm/science/description.html}};
\item Excluding phosphorescence spike events (5 events) \citep[e.g., see][]{Goldstein2019,WuBaiyangZhang2022}, which are listed in Table \ref{tab:burstlists}.

\end{enumerate}

\begin{figure}
    \includegraphics[width=\columnwidth]{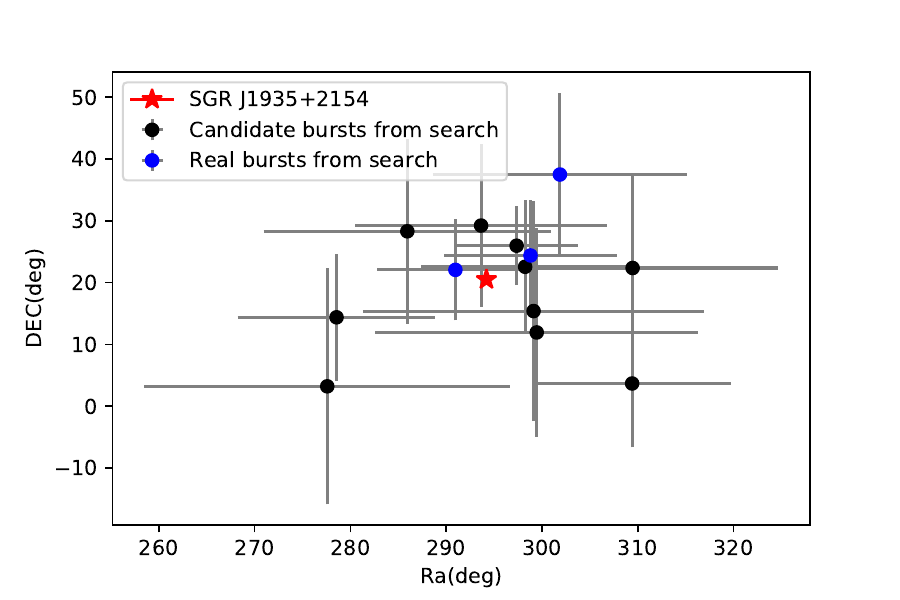}
    \vspace{-0.5cm}
    \caption{The location of candidate bursts. The red star marks the accurate position of SGR J1935+2154. The black and blue represent the location of candidate bursts and real (confirmed) bursts (i.e. which are also detected by Swift/BAT), respectively.  The gray lines mean 1-$\sigma$ confidence level.}
    \label{fig:loc}
\end{figure}

\begin{figure*}
    \includegraphics[width=\textwidth]{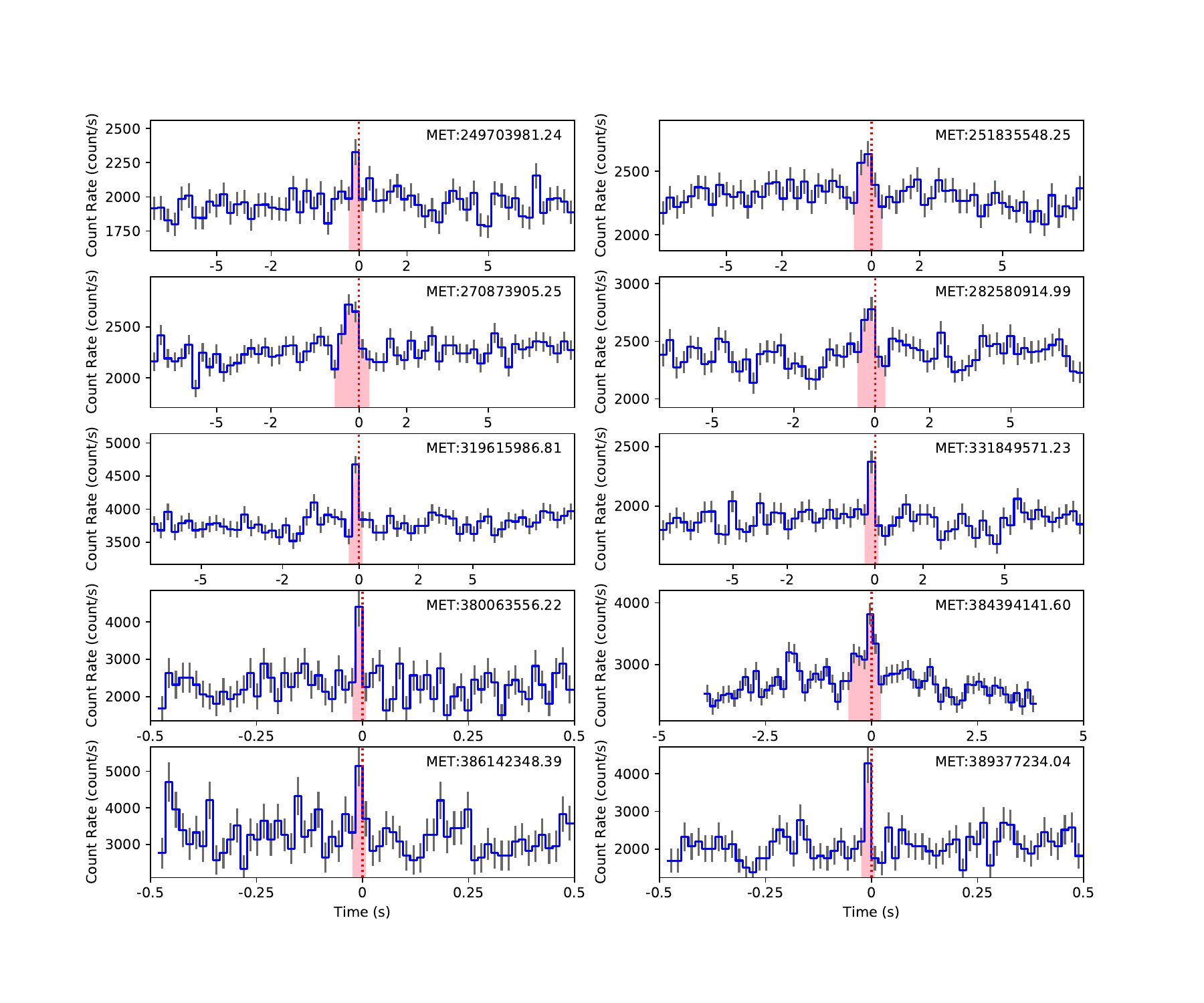}
    \vspace{-1.7cm}
    \caption{The light curve of candidate bursts from SGR J1935+2154. The blue curves represent the summed light curves of those NaI detectors with significant excess. The red dotted line represent the trigger time and pink shadow is the duration of candidate burst.}
    \label{fig:lc}
\end{figure*}

Finally, we found 10 new candidate bursts and 3 previously found bursts from SGR J1935+2154 during the time interval from 2008-08-12 to 2014-12-31 (UTC). We call these 10 bursts as candidates because they are solely detected by Fermi/GBM revealed by our targeted search and without confirmation by other instruments, while the other 3 bursts have been found and reported by Swift/BAT \citep[e.g.][]{Lin2020b}. Details of our burst candidates are listed in Table \ref{tab:burstlists}. The light curves and location of candidate bursts are shown in Fig. \ref{fig:lc} and Fig. \ref{fig:loc}, respectively. The location is generated by the search pipeline \citep{Cai2021}. Note that the light curve, spectrum and location of these candidate bursts are all consistent with known bursts from SGR J1935+2154, based on which we argue that most of these candidates, if not all, are probably from SGR J1935+2154.

\begin{table*}
\caption{The properties of SGR J1935+2154 burst candidates found by our targeted search in this work and the Phosphorescence Spikes Events.}
\label{tab:burstlists}
\begin{tabular}{lcccccccccc}
\hline
Event Time(UTC) & MaxLR $^\mathrm a$ & Significance & Template & Duration & RA       & DEC      & Err      & \multicolumn{3}{c}{BlackBody} \\ \cline{9-11} 
&    &    &    & (ms)   & ($^\circ$) & ($^\circ$) & ($^\circ$) & kT    & Flux$^\mathrm b$   & C-Stat/dof   \\ \hline
2008-11-30T02:13:00 & 22.16  &  6.74  & BlackBody & $<$256  &  285.94 & 28.3 & 14.99  & $3.69_{\mathrm -1.29}^{+2.57}$  &  $0.39_{\mathrm -0.16}^{+0.08}$ &  12.23/14 \\
2008-12-24T18:19:07 & 15.24  &  5.62  & BlackBody & $<$512  &  299.12 & 15.39 & 17.82  & $18.93_{\mathrm -5.77}^{+9.02}$ &  $0.77_{\mathrm -0.20}^{+0.23}$ &  12.23/14 \\
2009-08-02T02:45:03 & 18.56  &  6.17  & Powerlaw & $<$512  &  293.64 & 29.23 & 13.14  & $7.77_{\mathrm -3.21}^{+4.28}$  &  $0.59_{\mathrm -0.20}^{+0.16}$ &  15.38/14 \\
2009-12-15T14:41:52 & 21.6   &  6.61  & Powerlaw & $<$512  &  298.24 & 22.56 & 10.82  & $5.96_{\mathrm -4.43}^{+6.74}$  &  $0.34_{\mathrm -0.23}^{+0.02}$ &  22.52/14 \\
2011-02-17T06:13:04 & 49.18  &  10.05 & Comptonized  & $<$256  &  278.55 & 14.38 & 10.32  &  $8.23_{\mathrm -3.04}^{+4.21}$  & $0.63_{\mathrm -0.16}^{+0.11}$  & 21.53/22 \\
2011-07-08T20:26:09 & 18.95  &  5.72  & Powerlaw & $<$256  &  299.45 & 11.93 & 16.84  &  $1.39_{\mathrm -0.40}^{+0.40}$  & $0.70_{\mathrm -0.70}^{+0.04}$ &  9.73/14 \\
2013-01-16T21:12:33 & 20.73  &  6.59  & OTTB & 16  &  277.58 & 3.21 & 19.11  &  $7.23_{\mathrm -3.30}^{+4.59}$ & $1.47_{\mathrm -0.77}^{+2.14}$ & 18.92/30 \\
2013-03-08T00:08:58 & 46.33  &  6.64  & Powerlaw & 576  &  297.35 & 25.97 & 6.37  &  $4.40_{\mathrm -1.69}^{+2.48}$ & $1.04_{\mathrm -0.82}^{+1.15}$ & 23.9/30 \\
2013-03-28T05:45:45 & 23.17  &  6.93  & Powerlaw & 32  &  309.41 & 3.67 & 10.34  &  $4.09_{\mathrm -2.77}^{+1.84}$ & $0.85_{\mathrm -0.68}^{+0.28}$ & 7.83/14 \\
2013-05-04T16:20:31 & 21.32  &  6.75  & Powerlaw & 24  &  309.46 & 22.36 & 15.21  &  $7.54_{\mathrm -3.66}^{+7.65}$ & $1.35_{\mathrm -0.85}^{+1.95}$ & 12.78/30 \\
2014-07-05T09:32:48$^{\mathrm c}$ & 209.13 &  7.44  & OTTB & 96  &  298.8 & 24.39 & 9  &  $5.56_{\mathrm -1.28}^{+1.72}$ & $2.30_{\mathrm -2.01}^{+2.48}$ & 37.32/46 \\
2014-07-05T09:37:34$^{\mathrm c}$ & 291.28 &  21.37 & Powerlaw & 48  &  290.96 & 22.06 & 8.16  &  $6.26_{\mathrm -1.79}^{+2.36}$ & $2.48_{\mathrm -2.09}^{+2.86}$ & 46.73/46 \\
2014-07-05T09:41:06$^{\mathrm c}$ & 26.65  &  7.34  & OTTB & 64  &  301.87 & 37.49 & 13.26  &  $12.32_{\mathrm -3.74}^{+4.58}$ & $2.07_{\mathrm -1.40}^{+2.63}$ & 29.18/46 \\
\hline
\multicolumn{3}{l}{Phosphorescence Spikes Events}    &    &  &  &   &     &  &  &  \\ 
\hline
2009-01-28T15:32:06 & 60.08    & 12.22  & Powerlaw &  $<$256  &  302.46 & 12.22 & 3.75 &...&...&... \\
2010-02-10T13:34:14 & 317.86   & 22.73  & Powerlaw &  $<$256  &  287.39 & 13.19 & 2.56 &...&...&...\\
2010-12-16T06:03:10 & 37622.94 & 233.82 & Powerlaw &  $<$256  &  304.69 & 31.11 & 0.46 &...&...&...\\
2014-05-15T15:32:56 & 34.38    & 6.97   & BlackBody &  3  &  275.49 & 26.26 & 22.71 &...&...&...\\
2014-06-24T23:02:10 & 47.13    & 9.88   & Powerlaw &  7  &  281.95 & 5.69 & 16.97 &...&...&...\\
\hline
\multicolumn{11}{l}{\footnotesize$^\mathrm a$ Max Log-Likelihood Ratio \citep{Blackburn2015,Cai2021}}\\
\multicolumn{11}{l}{\footnotesize$^\mathrm b$ Values in units of $10^{-7} \ \mathrm{erg \ cm^{-2} \ s^{-1}}$ are calculated by using BlackBody Spectrum within 8-200 keV}\\
\multicolumn{11}{l}{\footnotesize$^\mathrm c$ Real burst which are also detected by Swift/BAT}\\
\end{tabular}
\end{table*}

%%%%%%%%%%%%%%%%%%%%%%%%%%%%%%%%%%%%%%%%%%%%%%%%%%

\begin{figure}
    \includegraphics[width=\columnwidth]{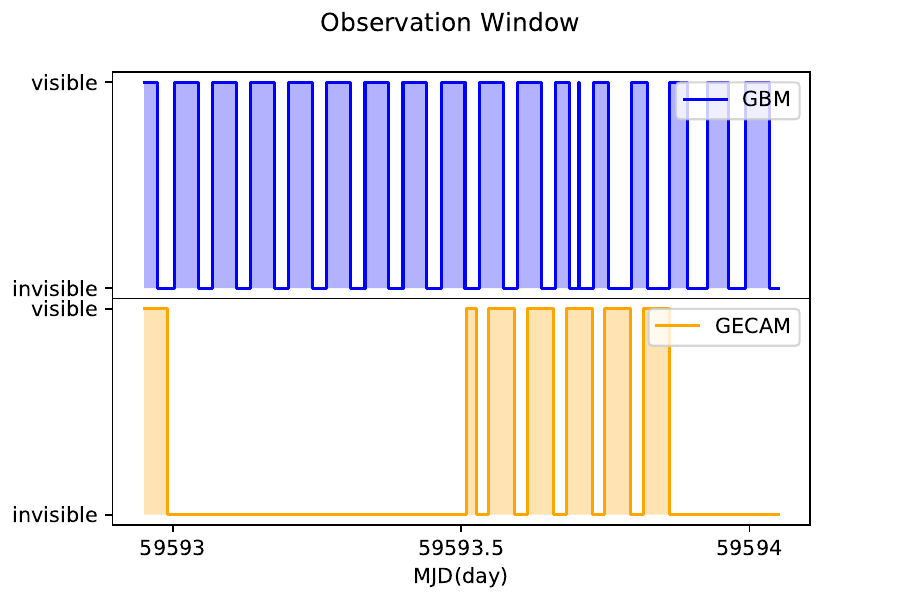}
    \vspace{-0.5cm}
    \caption{The visible time windows of Fermi/GBM and GECAM to SGR J1935+2154. The exposure time is the union of the Fermi/GBM and GECAM visible time interval in that day.}
    \label{fig:exposure}
\end{figure}

\begin{figure*}
    \includegraphics[width=1.0\textwidth]{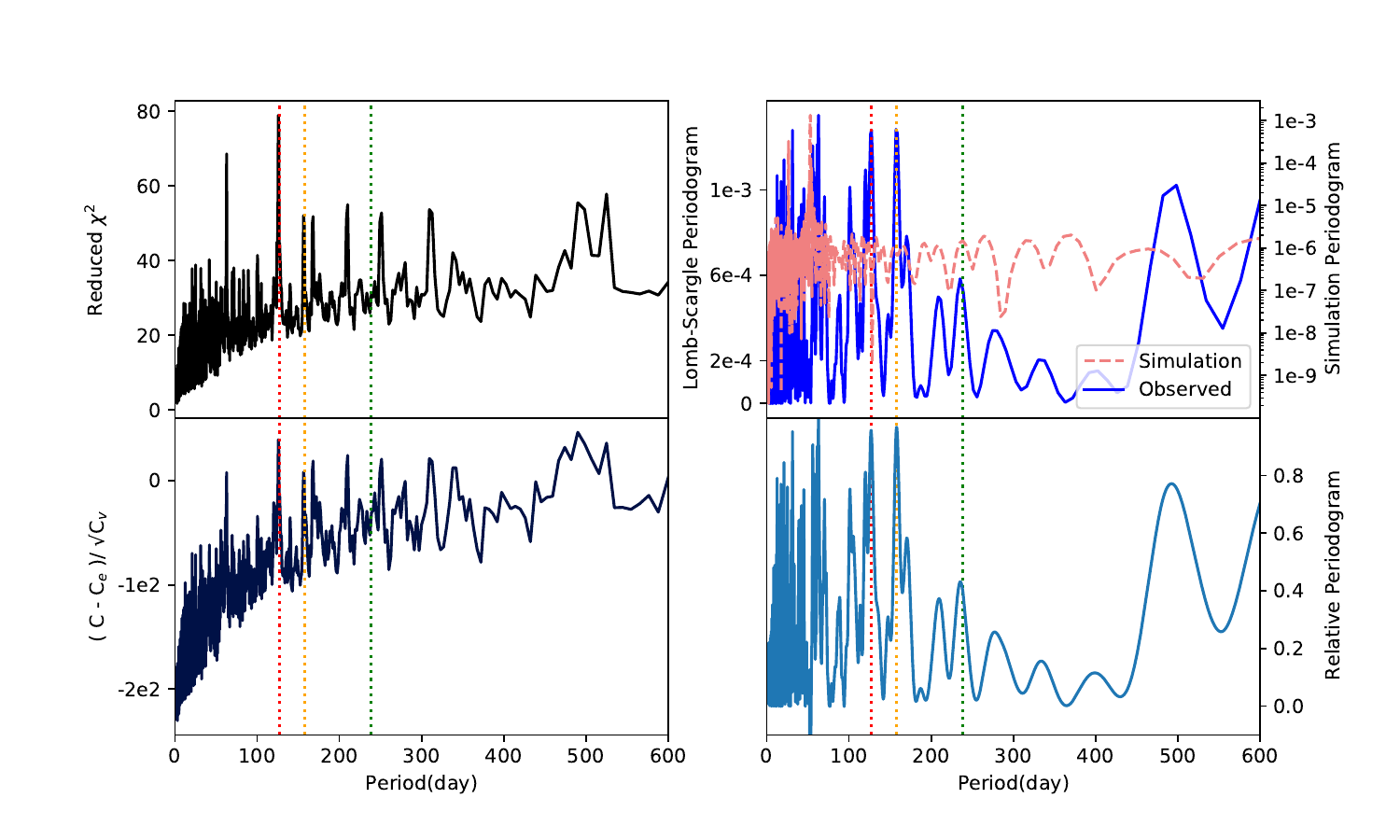}
    \vspace{-1.0cm}
    \caption{The period of SGR J1935+2154 derived from sample A. The left two panels show the result of reduced $ \chi^{2}$ and $(C-C_{\mathrm e})/\sqrt{C_{\mathrm v}}$ of period folding. The Lomb-Scargle periodogram of observed bursts data and simulated data of observation windows shows in the right upper panel with blue line and lightcoral dashed line, respectively. We normalize these two periodogram with maximum power value, and subtract the normalized periodogram of simulated data from the observed one. The result is shown in the right lower panel. The vertical red dotted line indicate the peak of relative periodogram which is 126.88 day, which is also significant in the period folding results with the reduced $ \chi^{2}$ and $(C-C_{\mathrm e})/\sqrt{C_{\mathrm v}}$ method. The vertical orange dotted and the vertical green dotted line represent 158.15 day and 238 day, respectively.}
    \label{fig:sampleA}
\end{figure*}

\begin{figure*}
    \includegraphics[width=1.0\textwidth]{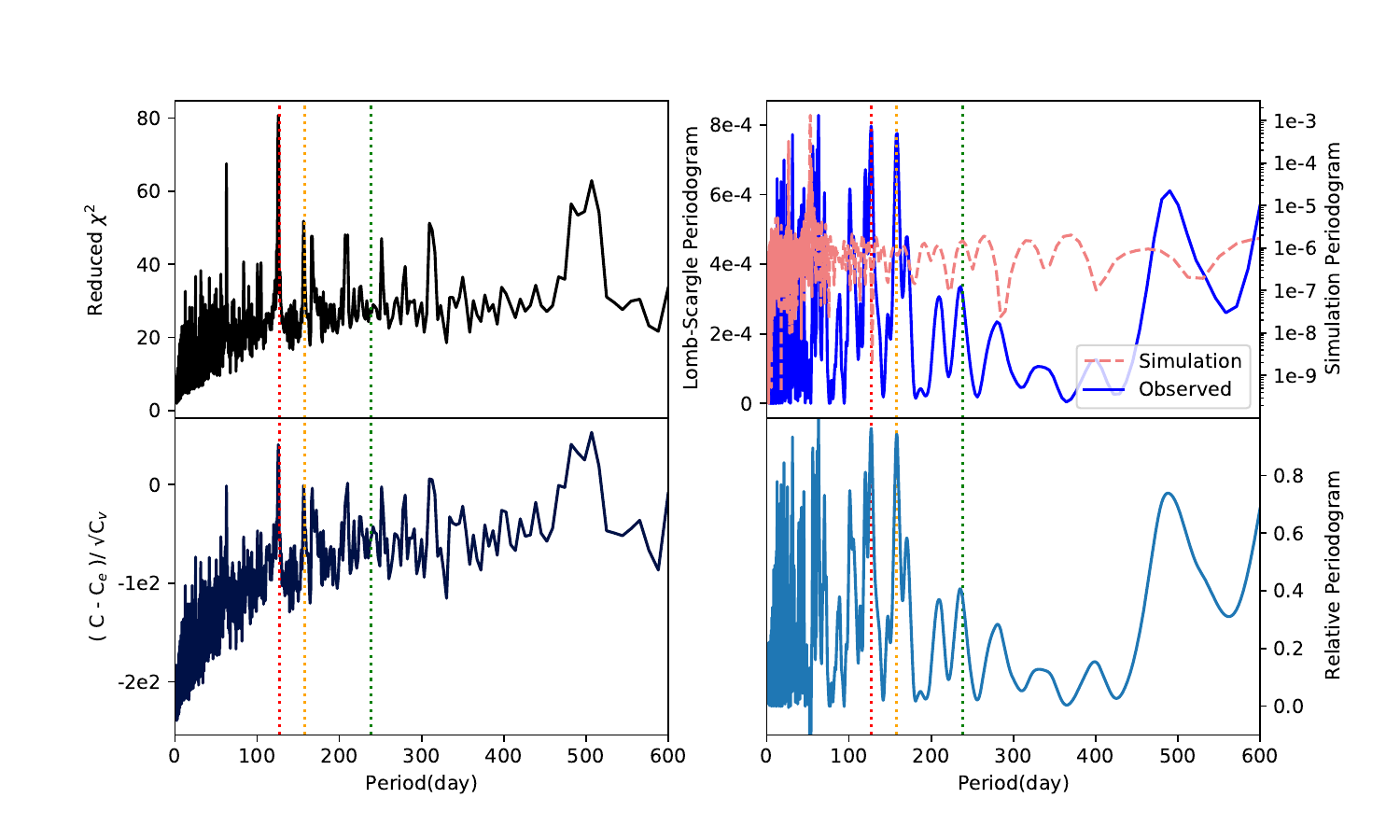}
    \vspace{-1.0cm}
    \caption{The period of SGR J1935+2154 derived from sample B. The left two panel show the result of reduced $ \chi^{2}$ and $(C-C_{\mathrm e})/\sqrt{C_{\mathrm v}}$ of period folding. The Lomb-Scargle periodogram of observed bursts data and simulated data of observation windows shows in the right upper panel with blue line and lightcoral dashed line, respectively. We normalize these two periodogram with maximum power value, and subtract normalized periodogram of simulated data from observed one. The result show in the right lower panel. The vertical red dotted line indicate the peak of relative periodogram which is 127.02 day. It is consistent with the peak of reduced $ \chi^{2}$ and $(C-C_{\mathrm e})/\sqrt{C_{\mathrm v}}$. The vertical orange dotted and the vertical green dotted line represent 158.15 day and 238 day, respectively}
    \label{fig:sampleB}
\end{figure*}

\begin{figure}
    \includegraphics[width=\columnwidth]{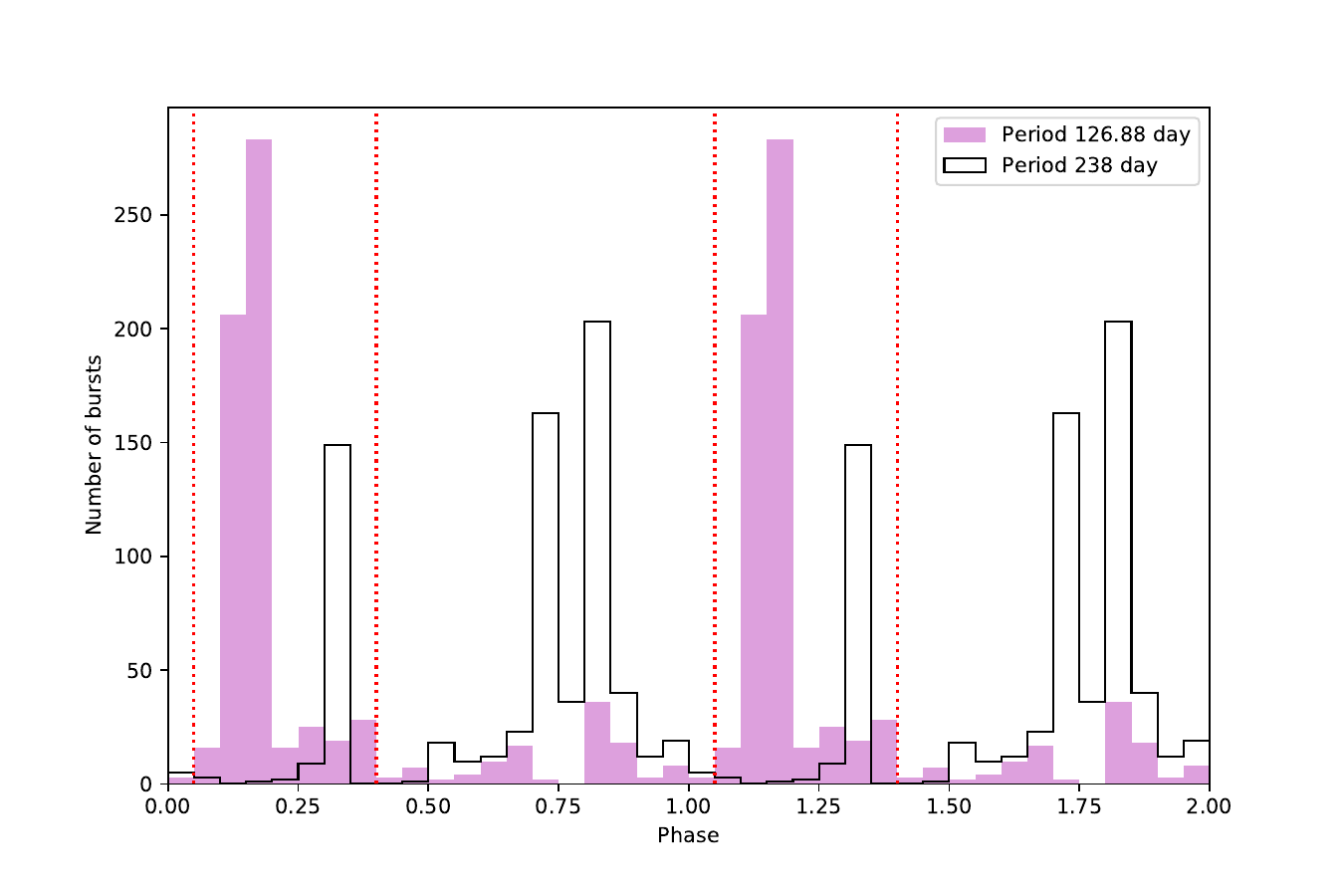}
    \vspace{-0.5cm}
    \caption{The folded phase of SGR J1935+2154. The plum histogram is the distribution of sample A with period 126.88 day. The active window (i.e. from $\varphi$=0.05 to $\varphi$=0.4) is denoted with red dotted lines. The black histogram represent the folding profile of Sample A with a period of 238 day \citep[reported in previous studies][]{Denissenya2021D, Grossan2021, Zou2021}, which is disfavored by the present results.}
    \label{fig:phase}
\end{figure}

\begin{figure}
    \includegraphics[width=\columnwidth]{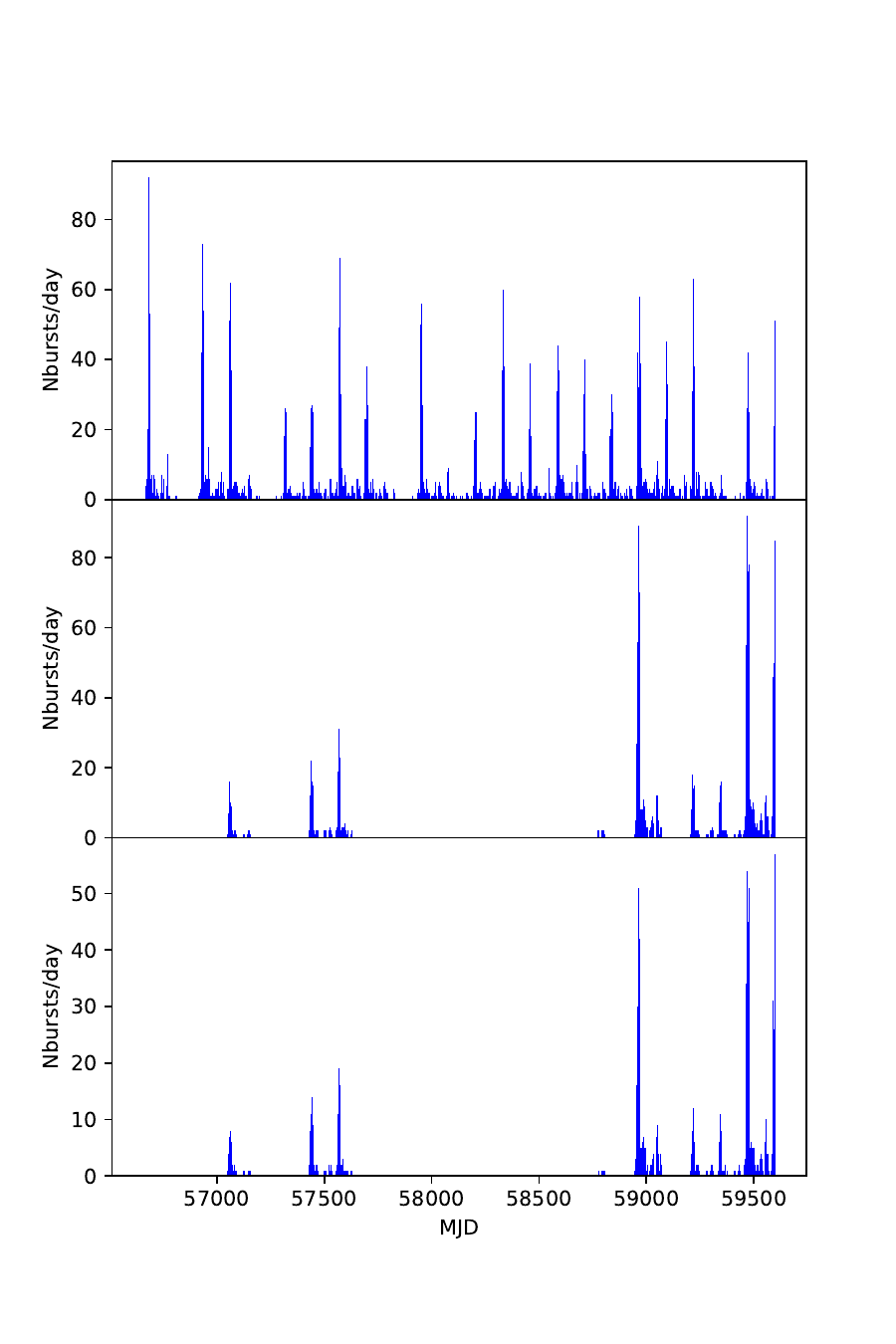}
    \vspace{-1.4cm}
    \caption{One of the 10 simulated burst data with an input period of 126.88 day as an example. The upper, middle and lower panel is the event rate of simulated bursts of Class 1, 2 and 3, respectively.}
    \label{fig:simu_bursts}
\end{figure}

\begin{figure*}
    \includegraphics[width=1.0\textwidth]{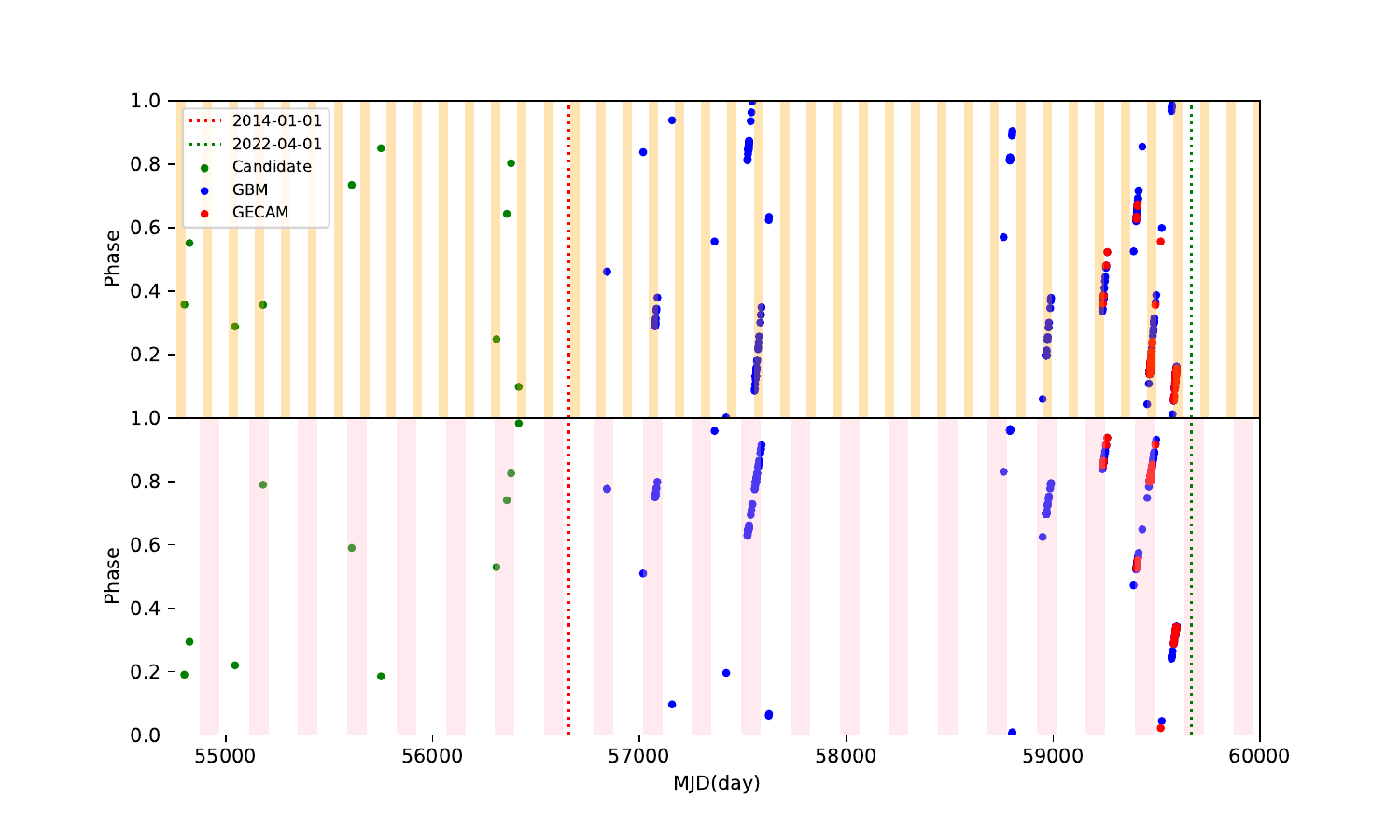}
    \vspace{-1.0cm}
    \caption{Upper panel: The burst times (MJD) and phase of SGR J1935+2154 with a period of 126.88 day. The orange shadow represent the active window in each period as shown in \ref{fig:phase}. The blue, red and green dots indicate the reported bursts of Fermi/GBM, GECAM and the targeted search candidate bursts of Fermi/GBM, respectively. Lower panel: The same with the upper panel but the period is set to 238 day and the pink shadow regions denote the active window \citep{Zou2021}.}
    \label{fig:burst_phase}
\end{figure*}

\begin{figure*}
    \includegraphics[width=1.0\textwidth]{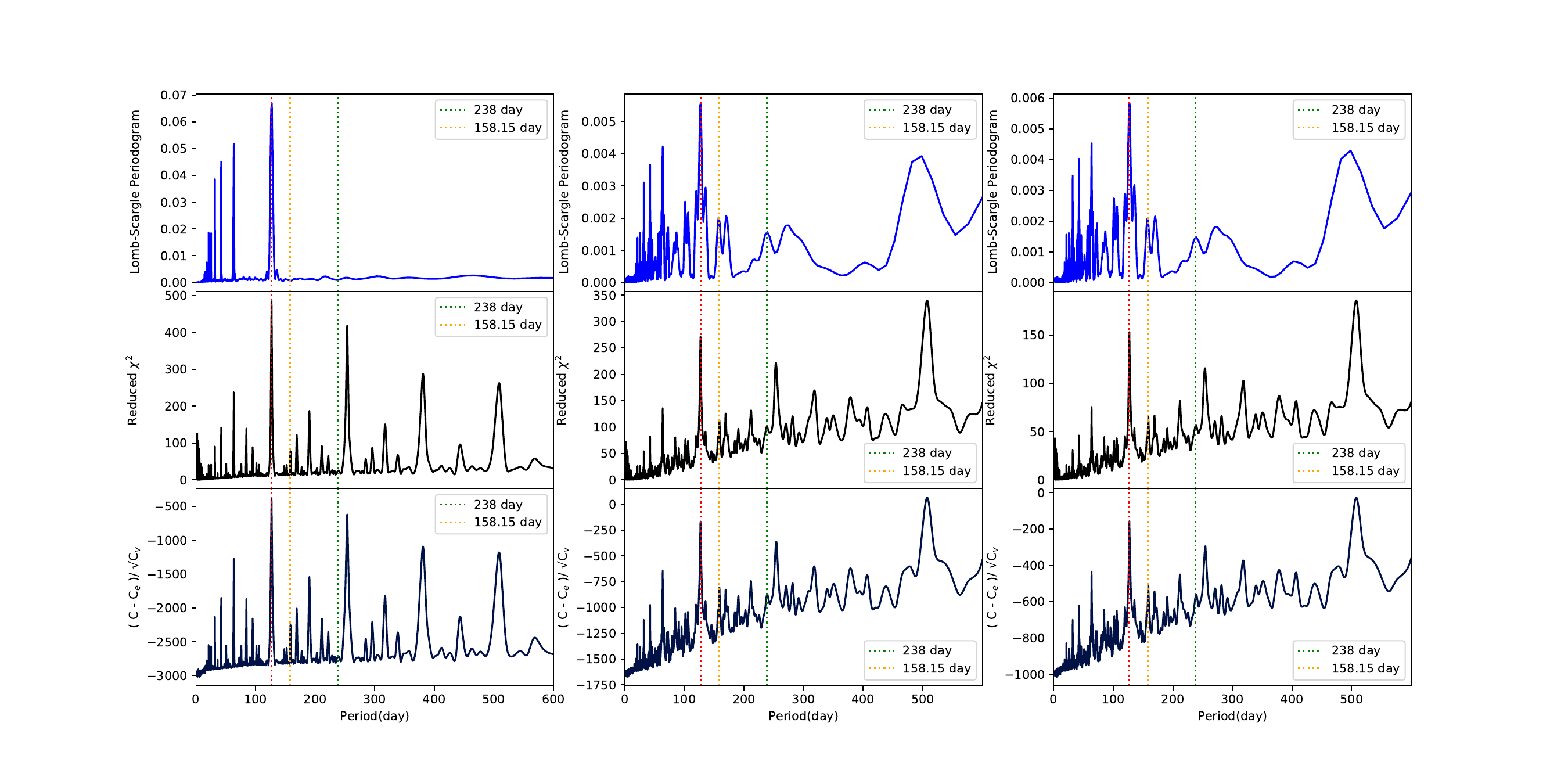}
    \vspace{-1.0cm}
    \caption{Period search on simulated bursts. The left, middle and right three panels are results for simulated burst data of Class 1, 2 and 3, respectively. The Lomb-Scargle periodogram and period folding with Pearson’s $\chi^{2}$-stat test and $C$-stat test results are shown from top to bottom row panels. The red dotted line represent 126.88 day, i.e. the input period for the simulation. The orange and green dotted lines represent 158.15 day and 238 day, which are the apparent periods of Lomb-Scargle periodogram of sample A.}
    \label{fig:simu_bursts_pt}
\end{figure*}

\section{Periodicity of Burst History}
\label{sec:pwb}

To explore the potential periodicity behavior in the burst history of SGR J1935+2154, we collected all bursts including the recently reported bursts (till 2022 January) from Fermi/GBM and GECAM, as well as the burst candidates we found in Fermi/GBM data from 2008-08-12 to 2014-12-31 (UTC). Fermi/GBM bursts that we used in this study have been reported in literature \citep[][]{Lin2020a,Lin2020b,Zou2021} or in the Gamma-ray Coordinates Network (GCN)\footnote{\url{https://gcn.gsfc.nasa.gov/}}. Since \cite{Zou2021} has listed the bursts of Fermi/GBM from 2014 July to 2021 October, we only listed Fermi/GBM bursts from 2021 November to 2022 January in Table \ref{tab:samples_gbm}. The burst sample of GECAM is listed in Table \ref{tab:samples_gecam}. 

We divide all bursts into two samples. Sample A contains all reported bursts from Fermi/GBM and GECAM from 2014 July to 2022 January. The start time (T0) of Sample A is set to 56658 (MJD). This is the main sample for searching the possible periodicity.
Sample B is composed of Sample A and candidate bursts found from 2008 to 2014. The start time of Sample B is set to 54771 (MJD). Owing to the small number of the candidate bursts, it has less constraint on the periodicity. Thus sample B is only used to check the periodicity result derived from Sample A. 

We note that the setting of the start time (T0) does not affect the periodicity result, which is confirmed by the simulations (see Section \ref{subsec:simu}). We also calculate the visible time windows and the exposure time history of SGR J1935+2154 for Fermi/GBM and GECAM respectively, which are used in the periodicity search. An example of the visible time window is shown in Fig \ref{fig:exposure}.

%%%%%%%%%%%%%%%%%%%%%%%%%%%%%%%%%%%%%%%%%%%%%%%%%%

\subsection{Period Folding}
\label{subsec:periodfold}

First, we adopt the Period Folding method to search the possible periodicity, which was used in the FRB studies \citep{CHIME2020a}. We fold the burst times of SGR J1935+2154 with different trial periods and group these folded bursts into $n$ bins, namely the number of phase bins. In this study, we set $n$ for period folding with both samples A $\&$ B. In addition, we also use the $C$-stat test together with Pearson’s $\chi^{2}$-stat test to examine the significance of the period ( with the folded burst numbers in phase bins). The $C$-stat can be written as 

\begin{equation}
C = 2\sum_{i=1}^{n} E_{i} - N_{i} +  N_{i} ln( N_{i} / E_{i} )
\label{equa:cstat}
\end{equation}
where $N_{\mathrm i}$ is the number of bursts in the phase bin $i$, $E_{\mathrm i}={p} {T}_{\mathrm i}$ is the expected number of bursts in bin ${i}$ if there is no period, ${T}_{\mathrm i}$ is the exposure time of bin ${i}$ and ${p}=\sum {N}_{\mathrm i}/\sum {T}_{\mathrm i}$ is the average burst rate. Like Reduced $\chi^{2}$, we use $({C}-{C}_{\mathrm {e}})/\sqrt{{C}_{\mathrm {v}}}$ to assess deviation and significance, where ${C}_{\mathrm {e}}$ is the expected value and ${C}_{\mathrm v}$ is variance \citep[see][]{Kaastra2017}. The peak of $({C}-{C}_{\mathrm {e}})/\sqrt{\ {C}_{\mathrm {v}}}$ indicate a possible period of bursts.

We set $n$ to 20 and search trial period from ${P}_{\mathrm {min}}$ = 0.5 days to ${P}_{\mathrm {max}}$ = 600 days ($f_{\mathrm {min}}=1/P_{\mathrm {max}}$, $ f_{\mathrm {max}}=1/P_{\mathrm {min}}$) with step $ \bigtriangleup f=0.1/T_{\mathrm {span}}$, where $ T_{\mathrm {span}}$ mean the longest time from the first burst to the last one. Setting $ P_{\mathrm{max}}=600$ to make sure that there are at least 5 periods in the extent of the whole observation of all bursts. The reduced $\chi^{2}$ and 
$(C-C_{\mathrm {e}})/\sqrt{C_{\mathrm {v}}}$ of two samples (Sample A and Sample B) are shown in the left two panels of Fig \ref{fig:sampleA} and \ref{fig:sampleB}, respectively. According to the results of sample A, the peak of reduced $ \chi^{2}$ is 127.31 day, which is the same as the peak of $ (C-C_{\mathrm {e}})/\sqrt{C_{\mathrm {v}}}$. The results of Period Folding are very consistent with that of periodogram method, which is 126.88 day, namely the vertical red dotted line in Fig \ref{fig:sampleA}. 

We note that, there is a peak around 55 days with similar significance. It should be caused by observation windows, according to the Lomb-Scargle Periodogram of simulation data of observation windows (see Section \ref{subsec:LS}). The peaks of $ \chi^{2}$ and $(C-C_{\mathrm {e}})/\sqrt{C_{\mathrm {v}}}$ of sample B are 127.86 day, respectively, which is well consistent with the result of sample A. 

%%%%%%%%%%%%%%%%%%%%%%%%%%%%%%%%%%%%%%%%%%%%%%%%%%

\subsection{Lomb-Scargle Periodogram}
\label{subsec:LS}

The Lomb-Scargle Periodogram has been widely used in unevenly sampled data for searching periodicity \citep{Lomb1976,Scargle1982,VanderPlas2018}. In this paper, we use the Lomb-Scargle function to search possible periodic window with test period spanning from 0.5 day to 600 day. The event rates mean that the burst numbers divided by exposure time and are calculated with a bin size of 0.05 day for stable False Alarm Probability (FAP) \citep[see][]{Zou2021}. The results are shown with blue line in right upper panel of Fig \ref{fig:sampleA} and \ref{fig:sampleB}, respectively. The most significant peak of the Lomb-Scargle periodogram is 127.30 day and 126.89 day for Sample A and B, respectively. We note that there are additional two non-negligible significant peaks 
around 157.2 day and 498.33 day. The FAP of each possible period are followings: FAP(127.30 day) = $3.39\times 10^{-13}$, FAP(157.2 day) = $4.76\times 10^{-13}$ and FAP(498.33 day) = $5.59\times 10^{-10}$. However, these two peaks are much less significant in the results of Period Folding method (left panels of Fig \ref{fig:sampleA} and \ref{fig:sampleB}). They are likely caused by observation windows and bursts gap, which will be discussed in Section \ref{subsec:simu}. 

We note that there is also a peak around 238 day, which is the favored period in previous studies \citep{Denissenya2021D, Grossan2021, Zou2021}, but the significance of this peak around 238 day is much lower than other peaks. This change of the significance of the period of 238 day is mostly caused by the inclusion of new bursts after 2021 October.

In addition, to test the effect of the observation window (see Fig. \ref{fig:exposure}), we simulate a series of bursts in visible observation windows of Fermi/GBM from 2014 July to 2022 January and calculate the Lomb-Scargle periodogram of these simulated bursts. The light coral dashed line represents the simulated data of observation windows in the right upper panel of Fig \ref{fig:sampleA} and \ref{fig:sampleB}. There are two apparent peaks at 26.46 day and 52.95 day, which indicate that the periods around 26 day or 52 day are probably caused by the observation window effects.

We normalize the periodogram of observation bursts data and the periodogram of simulated bursts considering observation window effect with maximum power value (namely the power of Lomb-Scargle periodogram), and subtract the latter from the former. Then we get the relative periodogram \citep{Zhang2021} which are shown in the right lower panels of Fig \ref{fig:sampleA} and \ref{fig:sampleB}. The results of the relative periodogram of two samples are similar to the original periodogram. The prominent peaks of the relative periodogram of sample A are 55, 126.88 and 158 day, which are consistent with that of sample B.

According to the above results, we conclude that the most favored period is about 126.88$\pm$2.05 day for Sample A and 127.02$\pm$1 day for Sample B. The error is calculated with the same method in \cite{CHIME2020a}: $ \sigma=PW_{\mathrm{active}}/T_{\mathrm{span}}$, where $ P$ is period and $ W_{\mathrm{active}}$ is active days. For sample A, we show the folded phase histogram in Fig \ref{fig:phase}. The filled plum histogram is the phase profile of sample A, based on which we define the phases between $\varphi$=0.05 and $\varphi$=0.4 as the active window. Then, we plot the time history (in MJD) of Fermi/GBM and GECAM bursts, together with active windows for periods of 126.88 day (this work) and 238 day \citep[reported in previous works,][]{Denissenya2021D, Grossan2021, Zou2021}, as shown in Fig \ref{fig:burst_phase}.
Most bursts fall in the active windows (see Fig \ref{fig:phase}), but some are outside, which are also shown in Fig \ref{fig:burst_phase}. This may hint that the period behavior is not very strict, e.g. including some quasi-periodic behavior.

%%%%%%%%%%%%%%%%%%%%%%%%%%%%%%%%%%%%%

\subsection{Simulation Tests}
\label{subsec:simu}

In this study, we test whether we can recover the input period in the burst series with the period folding and Lomb-Scargle periodogram methods used in this work. Especially we made two kinds of simulation tests to evaluate the effects of observation window and burst rate variation. 

First, we set the input period as $P_{\mathrm{simu}}$=126.88 day and then draw bursts according to the phase profile (see Fig \ref{fig:phase}) folded with this period of 126.88 day. Since we need to simulate bursts from 2014-01-01 to 2022-01-18, this time-span (8 years) contains about $\sim $23 periods. In each period, we make sure that the simulated burst rate follow the profile of folded phase.

Then, we did three different classes of simulations, with 10 simulations for each class. As an example, one of the 10 simulations with the input period of 126.88 day is shown in Fig \ref{fig:simu_bursts}). For class 1, the total number of simulated bursts in each period is random (upper panel in Fig \ref{fig:simu_bursts}). For class 2, the total number of bursts is equal to the detected bursts of SGR J1935+2154 (middle panel in Fig \ref{fig:simu_bursts}). For class 3, removing those bursts in the invisible time windows from the class 2. Then we apply the same period analysis with the Period Folding and Lomb-Scargle periodogram methods to the simulated bursts data.

We find that these 10 simulations give similar results for both Lomb-Scargle periodogram and period folding methods, thus we calculate the average Lomb-Scargle periodogram, $\chi^{2}$-stat and $C$-stat test of these 10 simulations. The results of all three classes of simulation are shown in Fig \ref{fig:simu_bursts_pt}. It shows that these periodicity search methods can successfully recover the input period (i.e. 126.88 day in this simulation). It also, as expected, shows several harmonic peaks of the input period. Interestingly, for class 2 and 3, both the periodogram and period folding show remarkable peaks around 498 day, while there are also lower significant peaks around the period of 158.15 day and 238 day. Since all these periods (i.e. 158.15, 238, and 498 day) are absent in the class 1, it seems that they are introduced by the burst rate variation or other observation effects.

Lastly, we repeat the above simulations with input period $P_{\mathrm{simu}}$=50, 100, 150, 200 and 250 day, respectively.
We find that the input period can be recovered by the periodicity search methods used in this study, which demonstrates that these methods of periodic analysis are reasonably reliable. Moreover, the Lomb-Scargle periodogram of some simulation periods (such as 50 day, 100 day, 150 day and 250 day) still show weak peaks around 125 day, 158 day, 238 day and 500 day. This indicates that the 126.88 day periodicity may be affected by observation factors as well.

%%%%%%%%%%%%%%%%%%%%%%%%%%%%%%%%%%%%%

\section{Discussions and Summary}
\label{sec:con}

The period behavior of repeating FRBs is still a mystery. Motivated by the connection between FRBs and SGRs, we analyze the possible active periodic window of bursts from SGR J1935+2154. In this paper, we adopt two methods to search the possible periodic window (Period Folding and Lomb-Scargle Periodogram) on the updated burst sample. 

To acquire more samples for searching and constraining the active periodic window, we developed a targeted search pipeline to search Fermi/GBM data from 2008 August to 2014 December. Then we exclude the solar flares, particle events, TGFs and some spurious triggers (such as phosphorescence spike events), and find 10 candidate bursts. Although these candidate bursts are found with Fermi/GBM data only and lack of confirmed observation by other instruments, their properties (light curve, spectrum and locations) are well consistent with known bursts from SGR J1935+2154, suggesting that most of them should be likely from SGR J1935+2154. They are rather weak that Fermi/GBM could not trigger them. None of them were in the field of view of Swift/BAT, thus there is no expected detection by Swift/BAT as well.

We defined two burst samples: sample A contains high confidence bursts reported by Fermi/GBM and GECAM (from 2014 July to 2022 January), while sample B is composed of sample A and these 10 candidate bursts found from Fermi/GBM data between the year of 2008 and 2014. We mainly use sample A to do the periodicity search and sample for the cross check the result of sample A.

Based on the period analysis of observation data and simulations, we find the most favored period is 126.88$\pm$2.05 day. The folded phase profile is shown in Fig \ref{fig:phase}, where the plum histogram is the phase for sample A, and the active window between $\varphi$=0.05 and $\varphi$=0.4 is given by Bayesian Block method. The burst history and phase are over-plotted in Fig \ref{fig:burst_phase} with the orange region representing the active window. It seems most of bursts lie in the active window including candidate bursts. There exist several peaks in the results of the searching periodic window. 
We implemented simulations to test how the observation effects may interfere the period analysis. Our simulations suggest that the period around 55 day is a fake signal which is likely caused by observation windows, and the periods around 158, 238 and 498 day could be reproduced if considering the burst rate variation among periods. 

Our results are very different from the periodicity $\sim$238 day of previous studies \citep{Grossan2021,Denissenya2021D,Zou2021}, which is mostly because different studies used different burst samples. In this study, we added about 200 new bursts found in a few months Fermi/GBM data and almost one year GECAM data. We also tried to extend the burst history before the first previously reported burst by Swift/BAT in 2014, by targeted search Fermi/GBM data from 2008 to 2014. Another difference is that we studied the observation effects on the periodicity search by detailed simulations and found that these effects may introduce fake periods.

However, we should note that, as shown in Fig. \ref{fig:burst_phase}, the periodicity 126.88 day is imperfect since the burst history of SGR J1935+2154 is very complicated and the burst rate varies drastically from time to time, even including some time epoch with no detection of any burst. This may hint that even though there is a period in the burst activity, the burst rate for some period epoch may be very low thus hardly to be observed.
At any rate, it seems any robust conclusion regarding the periodicity of the burst behavior should heavily reply on more observations in the future.

One of the popular explanations for the PWB is that the repeating FRBs come from a magnetar in binary with a companion star \citep{IokaZhang2020,Lyutikov2020,Zhang2020a,Wang2022}. The orbital motion of the binary leads to the periodic variation of the optical depth for the radio emission and, therefore, the radio emission can be observed only in a window during which the pulsar wind cavity is on the line of sight \citep{Chen2021}. However, there is actually no evidence supporting the existence of a companion star for SGR J1935+2154. Therefore, in principle, PWB may not be expected in the X-ray burst activity of SGR J1935+2154 if the PWB is primarily caused by the binary orbital motion. On the contrary, if the PWB indeed exists in SGR J1935+2154, then the PWB of the repeating FRBs might be also due to some other mechanisms \citep[e.g., the precession of the magnetar;][]{Levin2020,ZanazziLai2020,Chen2020,Wasserman2022} rather than the binary orbital motion.

If the period of 126.88 day is real, considering that SGR J1935+2154 is unlikely to be in a binary system, then we argue that one possible reason for this period behavior could be the free precession of neutron star \citep{ZanazziLai2020,Levin2020}. The period of the free precession may be caused by a strong internal magnetic field or deviation of the rotation axis of the magnetar. The precession period of a magnetar is given by \cite{Levin2020},
\begin{equation}
P_{\mathrm{prec}} \approx \frac{P_{\rm spin}}{\epsilon } \approx 20(\frac{k}{0.01}) ^{-1}\left(\frac{B_{\rm int}}{10^{16}  \rm G}\right)^{-2} \left(\frac{B_{\rm dip}}{10^{15}\rm G}\right) \left(\frac{t}{30  \rm yr} \right)^{1/2} \rm  days
\end{equation}
where $ P_{\mathrm{prec}}$ is the period of the free precession, $ P_{\mathrm{spin}}$ is the spin period of the magnetar, $ k$ is the numerical coefficient, $ B_{\mathrm{int}}$ is the internal magnetic field of the magnetar, $ B_{\mathrm{dip}}$ is the surface dipole magnetic field, and $ t$ is the age of the magnetar. The maximum $ k \approx 1$ would be approached if the field is fully coherent and purely toroidal. The value of $k$ is reduced when the field is tangled. According to this equation, using our result $ P_{\mathrm{prec}}=126.88$ day and $ P_{\mathrm{spin}} = 3.24$ s, $ B_{\mathrm{dip}}=2.2 \times 10^{14}$ G, $ t=3.6$ kyr \citep{Israel2016}, we can derive that $ \epsilon \approx 2.96 \times 10^{-7}$. If $ k=0.001$, the internal magnetic field is about $1.95 \times 10^{16}$ G. In addition, the cause for period could be the force precession as discussed by some studies \citep{Sob'yanin2020,Tong2020,YangZou2020}, e.g. the anomalous torque of electromagnetic forces induced by magnetar rotation could cause the precession; the torque caused by fallback disk \citep{Tong2020} can enhance the precession and lead to a long period.

\section*{acknowledgements}
This work is supported by the National Key R\&D Program of China (2021YFA0718500), the Strategic Priority Research Program on Space Science (Grant No. XDA15360102, XDA15360300, XDA15052700) of the Chinese Academy of Sciences, the National SKA program of China (2020SKA0120300), and the National Natural Science Foundation of China (Grant No. 11833003, 12173038).

\section*{Data Availability}
The Fermi/GBM data used in this paper are available at https://heasarc.gsfc.nasa.gov/FTP/fermi/data. The GECAM data will be released at http://gecam.ihep.ac.cn/. All data and codes underlying this article will be shared on reasonable request.

${Software}$:
Python\footnote{\url{https://www.python.org/}};
numpy \citep{harris2020array}\footnote{\url{https://numpy.org/}};
matplotlib \citep{Hunter2007}\footnote{\url{https://matplotlib.org/}};
scipy \citep{2020SciPy}\footnote{\url{https://scipy.org/}};
astropy \citep{Astropy2022}\footnote{\url{https://www.astropy.org/}};
GBM Data Tools \citep{GbmDataTools}\footnote{\url{https://fermi.gsfc.nasa.gov/ssc/data/analysis/rmfit/gbm_data_tools/gdt-docs/}};
Xspec \citep{Arnaud1996}\footnote{\url{https://heasarc.gsfc.nasa.gov/xanadu/xspec/}};

\begin{table*}
\caption{The SGR J1935+2154 burst sample detected by Fermi/GBM from 2021 November to 2022 January.}
\label{tab:samples_gbm}
\begin{tabular}{ccccc}
\hline
Burst Time (UTC) & Burst Time (UTC) & Burst Time (UTC) & Burst Time (UTC) & Burst Time (UTC) \\ 
\hline
2021-11-07T07:54:51.187 & 2021-12-24T03:42:34.341 & 2021-12-25T09:32:50.504 & 2021-12-25T18:48:28.387 & 2021-12-25T22:22:20.115 \\
2021-12-26T12:55:09.689 & 2021-12-29T16:41:26.191 & 2022-01-04T00:42:17.704 & 2022-01-04T04:32:11.147 & 2022-01-05T06:01:31.350 \\
2022-01-05T07:06:40.725 & 2022-01-06T02:36:14.044 & 2022-01-08T14:41:46.862 & 2022-01-09T04:57:16.080 & 2022-01-09T07:39:10.637 \\
2022-01-09T09:28:12.737 & 2022-01-09T09:40:33.491 & 2022-01-09T09:55:20.779 & 2022-01-09T12:57:21.913 & 2022-01-09T14:16:08.559 \\
2022-01-09T23:25:43.291 & 2022-01-10T02:57:16.776 & 2022-01-10T04:31:43.117 & 2022-01-10T15:55:38.291 & 2022-01-11T05:42:12.251 \\
2022-01-11T07:48:21.419 & 2022-01-11T08:58:35.308 & 2022-01-11T17:05:55.630 & 2022-01-11T18:21:07.672 & 2022-01-11T21:54:01.124 \\
2022-01-11T21:55:11.116 & 2022-01-12T01:03:46.329 & 2022-01-12T01:08:01.574 & 2022-01-12T02:05:58.212 & 2022-01-12T02:19:22.106 \\
2022-01-12T04:10:05.302 & 2022-01-12T04:31:19.934 & 2022-01-12T05:42:51.470 & 2022-01-12T08:39:25.279 & 2022-01-12T08:48:28.544 \\
2022-01-12T17:57:07.731 & 2022-01-12T18:12:04.642 & 2022-01-12T19:58:04.027 & 2022-01-12T23:00:12.748 & 2022-01-13T02:30:24.267 \\
2022-01-13T02:37:06.879 & 2022-01-13T05:59:14.025 & 2022-01-13T06:49:04.357 & 2022-01-13T08:34:20.010 & 2022-01-13T08:37:42.154 \\
2022-01-13T09:05:22.677 & 2022-01-13T10:28:58.992 & 2022-01-13T10:41:55.477 & 2022-01-13T13:07:30.634 & 2022-01-13T13:16:45.157 \\
2022-01-13T14:35:34.417 & 2022-01-13T14:40:54.853 & 2022-01-13T14:44:57.928 & 2022-01-13T14:58:01.149 & 2022-01-13T15:13:25.807 \\
2022-01-13T16:26:37.528 & 2022-01-13T19:23:10.210 & 2022-01-13T19:36:08.511 & 2022-01-13T19:57:55.715 & 2022-01-14T00:38:18.604 \\
2022-01-14T04:02:38.208 & 2022-01-14T06:37:56.739 & 2022-01-14T08:16:01.323 & 2022-01-14T11:26:42.754 & 2022-01-14T11:39:41.804 \\
2022-01-14T13:23:05.792 & 2022-01-14T13:34:30.163 & 2022-01-14T16:08:43.298 & 2022-01-14T19:27:16.671 & 2022-01-14T19:42:08.833 \\
2022-01-14T19:45:08.047 & 2022-01-14T20:41:14.905 & 2022-01-14T20:46:16.444 & 2022-01-14T20:46:17.510 & 2022-01-14T20:46:32.943 \\
2022-01-14T20:56:27.522 & 2022-01-14T22:21:29.360 & 2022-01-14T22:21:33.331 & 2022-01-14T22:38:26.261 & 2022-01-14T22:46:54.966 \\
2022-01-15T00:03:17.918 & 2022-01-15T00:36:54.124 & 2022-01-15T03:08:22.219 & 2022-01-15T06:37:08.499 & 2022-01-15T06:44:05.437 \\
2022-01-15T07:05:44.753 & 2022-01-15T07:08:43.476 & 2022-01-15T07:54:13.911 & 2022-01-15T07:54:43.936 & 2022-01-15T08:00:58.090 \\
2022-01-15T08:04:45.318 & 2022-01-15T08:16:31.251 & 2022-01-15T08:21:46.394 & 2022-01-15T08:23:02.089 & 2022-01-15T08:24:36.044 \\
2022-01-15T08:25:56.217 & 2022-01-15T08:27:38.893 & 2022-01-15T08:36:42.075 & 2022-01-15T08:36:51.475 & 2022-01-15T08:39:22.360 \\
2022-01-15T08:45:40.673 & 2022-01-15T09:24:40.334 & 2022-01-15T09:26:39.856 & 2022-01-15T09:39:04.523 & 2022-01-15T09:43:00.887 \\
2022-01-15T10:10:40.694 & 2022-01-15T12:50:12.834 & 2022-01-15T13:09:50.129 & 2022-01-15T13:13:05.353 & 2022-01-15T13:13:07.566 \\
2022-01-15T16:06:05.370 & 2022-01-15T16:10:49.654 & 2022-01-15T17:21:59.283 & 2022-01-15T19:11:28.185 & 2022-01-15T19:19:46.057 \\
2022-01-15T19:25:22.434 & 2022-01-15T19:27:50.194 & 2022-01-15T20:35:30.342 & 2022-01-15T22:38:44.484 & 2022-01-15T22:51:22.059 \\
2022-01-16T03:14:35.947 & 2022-01-16T07:43:05.577 & 2022-01-16T10:03:57.580 & 2022-01-16T10:48:37.617 & 2022-01-16T10:59:28.693 \\
2022-01-16T11:15:52.747 & 2022-01-16T11:17:35.433 & 2022-01-16T11:20:30.636 & 2022-01-16T11:30:23.385 & 2022-01-16T11:39:42.753 \\
2022-01-16T12:23:46.018 & 2022-01-16T12:38:23.736 & 2022-01-16T14:09:38.568 & 2022-01-16T14:10:01.110 & 2022-01-16T18:46:30.292 \\
2022-01-17T01:27:12.687 & 2022-01-17T01:39:37.260 & 2022-01-17T12:43:01.185 & 2022-01-17T15:37:58.722 & 2022-01-17T18:39:49.468 \\
\hline
\end{tabular}
\end{table*}

\begin{table*}
\caption{The SGR J1935+2154 burst sample detected by GECAM from 2021 January to 2022 January.}
\label{tab:samples_gecam}
\begin{tabular}{ccccc}
\hline
Burst Time (UTC) & Burst Time (UTC) & Burst Time (UTC) & Burst Time (UTC) & Burst Time (UTC) \\ 
\hline
2021-01-27T06:50:20.750 & 2021-01-30T08:39:53.840 & 2021-01-30T10:35:35.120 & 2021-02-11T13:43:16.760 & 2021-02-16T22:20:39.600 \\
2021-07-07T00:33:31.640 & 2021-07-08T00:18:18.560 & 2021-07-12T04:32:39.600 & 2021-07-12T22:12:58.100 & 2021-09-09T21:07:12.150 \\
2021-09-10T01:04:33.500 & 2021-09-10T02:07:56.700 & 2021-09-10T02:08:28.800 & 2021-09-10T03:22:40.550 & 2021-09-10T03:24:47.150 \\
2021-09-10T03:42:45.750 & 2021-09-10T05:05:03.350 & 2021-09-10T05:35:55.500 & 2021-09-11T16:35:46.500 & 2021-09-11T16:39:21.000 \\
2021-09-11T16:50:03.850 & 2021-09-11T17:01:10.800 & 2021-09-11T17:01:59.550 & 2021-09-11T17:04:29.800 & 2021-09-11T17:10:48.750 \\
2021-09-11T18:02:13.500 & 2021-09-11T18:04:46.350 & 2021-09-11T18:54:36.050 & 2021-09-11T19:43:28.000 & 2021-09-11T19:46:50.050 \\
2021-09-11T20:13:40.550 & 2021-09-11T20:22:59.050 & 2021-09-11T20:33:14.550 & 2021-09-11T21:07:28.350 & 2021-09-11T22:51:41.600 \\
2021-09-12T00:34:37.450 & 2021-09-12T00:45:49.400 & 2021-09-12T05:14:07.950 & 2021-09-12T05:44:17.050 & 2021-09-12T16:26:08.150 \\
2021-09-12T16:52:07.950 & 2021-09-12T22:16:36.200 & 2021-09-13T00:27:25.200 & 2021-09-13T14:12:39.650 & 2021-09-13T19:51:33.350 \\
2021-09-14T11:10:36.250 & 2021-09-14T14:15:42.900 & 2021-09-14T23:21:58.500 & 2021-09-14T23:26:34.050 & 2021-09-15T02:39:25.700 \\
2021-09-15T15:32:56.050 & 2021-09-17T12:52:37.800 & 2021-09-17T13:58:25.100 & 2021-09-18T22:58:52.150 & 2021-09-22T02:39:10.200 \\
2021-09-22T20:12:16.500 & 2021-10-07T11:57:07.700 & 2021-11-01T23:13:41.950 & 2022-01-04T04:32:11.200 & 2022-01-05T06:01:31.450 \\
2022-01-05T07:06:40.800 & 2022-01-06T02:36:14.100 & 2022-01-08T14:41:46.900 & 2022-01-09T07:39:10.700 & 2022-01-10T06:52:40.500 \\
2022-01-11T08:58:35.450 & 2022-01-12T01:03:46.900 & 2022-01-12T05:42:51.650 & 2022-01-12T08:39:25.450 & 2022-01-12T17:57:08.500 \\
2022-01-13T19:36:08.600 & 2022-01-13T20:14:58.600 & 2022-01-13T21:41:17.900 & 2022-01-14T19:42:08.833 & 2022-01-14T19:45:08.100 \\
2022-01-14T19:56:52.700 & 2022-01-14T20:06:07.400 & 2022-01-14T20:07:03.050 & 2022-01-14T20:12:45.300 & 2022-01-14T20:15:54.400 \\
2022-01-14T20:21:05.150 & 2022-01-14T20:23:35.400 & 2022-01-14T20:26:50.300 & 2022-01-14T20:29:07.250 & 2022-01-14T20:31:49.900 \\
2022-01-15T09:26:39.900 & 2022-01-15T13:52:26.050 & 2022-01-15T16:31:14.900 & 2022-01-15T17:21:59.300 & 2022-01-16T10:48:37.650 \\
2022-01-17T01:39:37.300 & 2022-01-23T20:06:38.750 & 2022-01-24T02:10:55.050 & 2022-01-24T02:27:07.400 &                         \\
\hline
\end{tabular}
\end{table*}



\bibliographystyle{mnras}
\bibliography{main} % if your bibtex file is called example.bib

@ARTICLE{Zou2021,
       author = {{Zou}, Jin-Hang and {Zhang}, Bin-Bin and {Zhang}, Guo-Qiang and {Yang}, Yu-Han and {Shao}, Lang and {Wang}, Fa-Yin},
        title = "{Periodicity Search on X-Ray Bursts of SGR J1935+2154 Using 8.5 yr of Fermi/GBM Data}",
      journal = {\apjl},
         year = 2021,
        month = dec,
       volume = {923},
       number = {2},
          eid = {L30},
        pages = {L30},
          doi = {10.3847/2041-8213/ac3759},
archivePrefix = {arXiv},
       eprint = {2107.03800},
 primaryClass = {astro-ph.HE},
       adsurl = {https://ui.adsabs.harvard.edu/abs/2021ApJ...923L..30Z}
}

@ARTICLE{Kouveliotou1998,
       author = {{Kouveliotou}, C. and {Dieters}, S. and {Strohmayer}, T. and {van Paradijs}, J. and {Fishman}, G.~J. and {Meegan}, C.~A. and {Hurley}, K. and {Kommers}, J. and {Smith}, I. and {Frail}, D. and {Murakami}, T.},
        title = "{An X-ray pulsar with a superstrong magnetic field in the soft {\ensuremath{\gamma}}-ray repeater SGR1806 - 20}",
      journal = {\nat},
         year = 1998,
        month = may,
       volume = {393},
       number = {6682},
        pages = {235-237},
          doi = {10.1038/30410},
       adsurl = {https://ui.adsabs.harvard.edu/abs/1998Natur.393..235K}
}

@ARTICLE{vanKerkwijk1995,
       author = {{van Kerkwijk}, M.~H. and {Kulkarni}, S.~R. and {Matthews}, K. and {Neugebauer}, G.},
        title = "{A Luminous Companion to SGR 1806-20}",
      journal = {\apjl},
         year = 1995,
        month = may,
       volume = {444},
        pages = {L33},
          doi = {10.1086/187853},
archivePrefix = {arXiv},
       eprint = {astro-ph/9502038},
 primaryClass = {astro-ph},
       adsurl = {https://ui.adsabs.harvard.edu/abs/1995ApJ...444L..33V}
}

@ARTICLE{Banas1997,
       author = {{Banas}, Kenneth R. and {Hughes}, John P. and {Bronfman}, L. and {Nyman}, L. -{\r{A}}.},
        title = "{Supernova Remnants Associated with Molecular Clouds in the Large Magellanic Cloud}",
      journal = {\apj},
         year = 1997,
        month = may,
       volume = {480},
       number = {2},
        pages = {607-617},
          doi = {10.1086/303989},
archivePrefix = {arXiv},
       eprint = {astro-ph/9612198},
 primaryClass = {astro-ph},
       adsurl = {https://ui.adsabs.harvard.edu/abs/1997ApJ...480..607B}
}

@ARTICLE{KaspiBeloborodov2017,
       author = {{Kaspi}, Victoria M. and {Beloborodov}, Andrei M.},
        title = "{Magnetars}",
      journal = {\araa},
         year = 2017,
        month = aug,
       volume = {55},
       number = {1},
        pages = {261-301},
          doi = {10.1146/annurev-astro-081915-023329},
archivePrefix = {arXiv},
       eprint = {1703.00068},
 primaryClass = {astro-ph.HE},
       adsurl = {https://ui.adsabs.harvard.edu/abs/2017ARA&A..55..261K}
}

@ARTICLE{DuncanThompson1992,
       author = {{Duncan}, Robert C. and {Thompson}, Christopher},
        title = "{Formation of Very Strongly Magnetized Neutron Stars: Implications for Gamma-Ray Bursts}",
      journal = {\apjl},
         year = 1992,
        month = jun,
       volume = {392},
        pages = {L9},
          doi = {10.1086/186413},
       adsurl = {https://ui.adsabs.harvard.edu/abs/1992ApJ...392L...9D}
}

@article{Lorimer2007,
author = {D. R. Lorimer  and M. Bailes  and M. A. McLaughlin  and D. J. Narkevic  and F. Crawford },
title = {A Bright Millisecond Radio Burst of Extragalactic Origin},
journal = {Science},
volume = {318},
number = {5851},
pages = {777-780},
year = {2007},
doi = {10.1126/science.1147532},
URL = {https://www.science.org/doi/abs/10.1126/science.1147532},
eprint = {https://www.science.org/doi/pdf/10.1126/science.1147532}}

@article{Thornton2013,
author = {D. Thornton  and B. Stappers  and M. Bailes  and B. Barsdell  and S. Bates  and N. D. R. Bhat  and M. Burgay  and S. Burke-Spolaor  and D. J. Champion  and P. Coster  and N. D'Amico  and A. Jameson  and S. Johnston  and M. Keith  and M. Kramer  and L. Levin  and S. Milia  and C. Ng  and A. Possenti  and W. van Straten },
title = {A Population of Fast Radio Bursts at Cosmological Distances},
journal = {Science},
volume = {341},
number = {6141},
pages = {53-56},
year = {2013},
doi = {10.1126/science.1236789},
URL = {https://www.science.org/doi/abs/10.1126/science.1236789},
eprint = {https://www.science.org/doi/pdf/10.1126/science.1236789}}

@ARTICLE{Tendulkar2017,
       author = {{Tendulkar}, S.~P. and {Bassa}, C.~G. and {Cordes}, J.~M. and {Bower}, G.~C. and {Law}, C.~J. and {Chatterjee}, S. and {Adams}, E.~A.~K. and {Bogdanov}, S. and {Burke-Spolaor}, S. and {Butler}, B.~J. and {Demorest}, P. and {Hessels}, J.~W.~T. and {Kaspi}, V.~M. and {Lazio}, T.~J.~W. and {Maddox}, N. and {Marcote}, B. and {McLaughlin}, M.~A. and {Paragi}, Z. and {Ransom}, S.~M. and {Scholz}, P. and {Seymour}, A. and {Spitler}, L.~G. and {van Langevelde}, H.~J. and {Wharton}, R.~S.},
        title = "{The Host Galaxy and Redshift of the Repeating Fast Radio Burst FRB 121102}",
      journal = {\apjl},
         year = 2017,
        month = jan,
       volume = {834},
       number = {2},
          eid = {L7},
        pages = {L7},
          doi = {10.3847/2041-8213/834/2/L7},
archivePrefix = {arXiv},
       eprint = {1701.01100},
 primaryClass = {astro-ph.HE},
       adsurl = {https://ui.adsabs.harvard.edu/abs/2017ApJ...834L...7T}
}

@ARTICLE{Kulkarni2014,
       author = {{Kulkarni}, S.~R. and {Ofek}, E.~O. and {Neill}, J.~D. and {Zheng}, Z. and {Juric}, M.},
        title = "{Giant Sparks at Cosmological Distances?}",
      journal = {\apj},
         year = 2014,
        month = dec,
       volume = {797},
       number = {1},
          eid = {70},
        pages = {70},
          doi = {10.1088/0004-637X/797/1/70},
archivePrefix = {arXiv},
       eprint = {1402.4766},
 primaryClass = {astro-ph.HE},
       adsurl = {https://ui.adsabs.harvard.edu/abs/2014ApJ...797...70K}
}

@ARTICLE{CHIME2020a,
       author = {{CHIME/FRB Collaboration} and {Amiri}, M. and {Andersen}, B.~C. and {Bandura}, K.~M. and {Bhardwaj}, M. and {Boyle}, P.~J. and {Brar}, C. and {Chawla}, P. and {Chen}, T. and {Cliche}, J.~F. and {Cubranic}, D. and {Deng}, M. and {Denman}, N.~T. and {Dobbs}, M. and {Dong}, F.~Q. and {Fandino}, M. and {Fonseca}, E. and {Gaensler}, B.~M. and {Giri}, U. and {Good}, D.~C. and {Halpern}, M. and {Hessels}, J.~W.~T. and {Hill}, A.~S. and {H{\"o}fer}, C. and {Josephy}, A. and {Kania}, J.~W. and {Karuppusamy}, R. and {Kaspi}, V.~M. and {Keimpema}, A. and {Kirsten}, F. and {Landecker}, T.~L. and {Lang}, D.~A. and {Leung}, C. and {Li}, D.~Z. and {Lin}, H. -H. and {Marcote}, B. and {Masui}, K.~W. and {McKinven}, R. and {Mena-Parra}, J. and {Merryfield}, M. and {Michilli}, D. and {Milutinovic}, N. and {Mirhosseini}, A. and {Naidu}, A. and {Newburgh}, L.~B. and {Ng}, C. and {Nimmo}, K. and {Paragi}, Z. and {Patel}, C. and {Pen}, U. -L. and {Pinsonneault-Marotte}, T. and {Pleunis}, Z. and {Rafiei-Ravandi}, M. and {Rahman}, M. and {Ransom}, S.~M. and {Renard}, A. and {Sanghavi}, P. and {Scholz}, P. and {Shaw}, J.~R. and {Shin}, K. and {Siegel}, S.~R. and {Singh}, S. and {Smegal}, R.~J. and {Smith}, K.~M. and {Stairs}, I.~H. and {Tendulkar}, S.~P. and {Tretyakov}, I. and {Vanderlinde}, K. and {Wang}, H. and {Wang}, X. and {Wulf}, D. and {Yadav}, P. and {Zwaniga}, A.~V.},
        title = "{Periodic activity from a fast radio burst source}",
      journal = {\nat},
         year = 2020,
        month = jun,
       volume = {582},
       number = {7812},
        pages = {351-355},
          doi = {10.1038/s41586-020-2398-2},
archivePrefix = {arXiv},
       eprint = {2001.10275},
 primaryClass = {astro-ph.HE},
       adsurl = {https://ui.adsabs.harvard.edu/abs/2020Natur.582..351C}
}

@ARTICLE{CHIME2020b,
       author = {{CHIME/FRB Collaboration} and {Andersen}, B.~C. and {Bandura}, K.~M. and {Bhardwaj}, M. and {Bij}, A. and {Boyce}, M.~M. and {Boyle}, P.~J. and {Brar}, C. and {Cassanelli}, T. and {Chawla}, P. and {Chen}, T. and {Cliche}, J. -F. and {Cook}, A. and {Cubranic}, D. and {Curtin}, A.~P. and {Denman}, N.~T. and {Dobbs}, M. and {Dong}, F.~Q. and {Fandino}, M. and {Fonseca}, E. and {Gaensler}, B.~M. and {Giri}, U. and {Good}, D.~C. and {Halpern}, M. and {Hill}, A.~S. and {Hinshaw}, G.~F. and {H{\"o}fer}, C. and {Josephy}, A. and {Kania}, J.~W. and {Kaspi}, V.~M. and {Landecker}, T.~L. and {Leung}, C. and {Li}, D.~Z. and {Lin}, H. -H. and {Masui}, K.~W. and {McKinven}, R. and {Mena-Parra}, J. and {Merryfield}, M. and {Meyers}, B.~W. and {Michilli}, D. and {Milutinovic}, N. and {Mirhosseini}, A. and {M{\"u}nchmeyer}, M. and {Naidu}, A. and {Newburgh}, L.~B. and {Ng}, C. and {Patel}, C. and {Pen}, U. -L. and {Pinsonneault-Marotte}, T. and {Pleunis}, Z. and {Quine}, B.~M. and {Rafiei-Ravandi}, M. and {Rahman}, M. and {Ransom}, S.~M. and {Renard}, A. and {Sanghavi}, P. and {Scholz}, P. and {Shaw}, J.~R. and {Shin}, K. and {Siegel}, S.~R. and {Singh}, S. and {Smegal}, R.~J. and {Smith}, K.~M. and {Stairs}, I.~H. and {Tan}, C.~M. and {Tendulkar}, S.~P. and {Tretyakov}, I. and {Vanderlinde}, K. and {Wang}, H. and {Wulf}, D. and {Zwaniga}, A.~V.},
        title = "{A bright millisecond-duration radio burst from a Galactic magnetar}",
      journal = {\nat},
         year = 2020,
        month = nov,
       volume = {587},
       number = {7832},
        pages = {54-58},
          doi = {10.1038/s41586-020-2863-y},
archivePrefix = {arXiv},
       eprint = {2005.10324},
 primaryClass = {astro-ph.HE},
       adsurl = {https://ui.adsabs.harvard.edu/abs/2020Natur.587...54C}
}

@ARTICLE{Yang2021,
       author = {{Yang}, Yu-Han and {Zhang}, Bin-Bin and {Lin}, Lin and {Zhang}, Bing and {Zhang}, Guo-Qiang and {Yang}, Yi-Si and {Tu}, Zuo-Lin and {Zou}, Jin-Hang and {Ye}, Hao-Yang and {Wang}, Fa-Yin and {Dai}, Zi-Gao},
        title = "{Bursts before Burst: A Comparative Study on FRB 200428-associated and FRB-absent X-Ray Bursts from SGR J1935+2154}",
      journal = {\apjl},
         year = 2021,
        month = jan,
       volume = {906},
       number = {2},
          eid = {L12},
        pages = {L12},
          doi = {10.3847/2041-8213/abd02a},
archivePrefix = {arXiv},
       eprint = {2009.10342},
 primaryClass = {astro-ph.HE},
       adsurl = {https://ui.adsabs.harvard.edu/abs/2021ApJ...906L..12Y}
}

@ARTICLE{Rajwade2020,
       author = {{Rajwade}, K.~M. and {Mickaliger}, M.~B. and {Stappers}, B.~W. and {Morello}, V. and {Agarwal}, D. and {Bassa}, C.~G. and {Breton}, R.~P. and {Caleb}, M. and {Karastergiou}, A. and {Keane}, E.~F. and {Lorimer}, D.~R.},
        title = "{Possible periodic activity in the repeating FRB 121102}",
      journal = {\mnras},
         year = 2020,
        month = jul,
       volume = {495},
       number = {4},
        pages = {3551-3558},
          doi = {10.1093/mnras/staa1237},
archivePrefix = {arXiv},
       eprint = {2003.03596},
 primaryClass = {astro-ph.HE},
       adsurl = {https://ui.adsabs.harvard.edu/abs/2020MNRAS.495.3551R}
}

@ARTICLE{Cruces2021,
       author = {{Cruces}, M. and {Spitler}, L.~G. and {Scholz}, P. and {Lynch}, R. and {Seymour}, A. and {Hessels}, J.~W.~T. and {Gouiff{\'e}s}, C. and {Hilmarsson}, G.~H. and {Kramer}, M. and {Munjal}, S.},
        title = "{Repeating behaviour of FRB 121102: periodicity, waiting times, and energy distribution}",
      journal = {\mnras},
         year = 2021,
        month = jan,
       volume = {500},
       number = {1},
        pages = {448-463},
          doi = {10.1093/mnras/staa3223},
archivePrefix = {arXiv},
       eprint = {2008.03461},
 primaryClass = {astro-ph.HE},
       adsurl = {https://ui.adsabs.harvard.edu/abs/2021MNRAS.500..448C}
}

@ARTICLE{Grossan2021,
       author = {{Grossan}, Bruce},
        title = "{Possible Periodic Windowed Behavior in SGR1935+2154 Bursts}",
      journal = {\pasp},
         year = 2021,
        month = jul,
       volume = {133},
       number = {1025},
          eid = {074202},
        pages = {074202},
          doi = {10.1088/1538-3873/ac07b1},
archivePrefix = {arXiv},
       eprint = {2006.16480},
 primaryClass = {astro-ph.HE},
       adsurl = {https://ui.adsabs.harvard.edu/abs/2021PASP..133g4202G}
}

@ARTICLE{Denissenya2021D,
       author = {{Denissenya}, Mikhail and {Grossan}, Bruce and {Linder}, Eric V.},
        title = "{Distinguishing time clustering of astrophysical bursts}",
      journal = {\prd},
         year = 2021,
        month = jul,
       volume = {104},
       number = {2},
          eid = {023007},
        pages = {023007},
          doi = {10.1103/PhysRevD.104.023007},
archivePrefix = {arXiv},
       eprint = {2103.10618},
 primaryClass = {astro-ph.IM},
       adsurl = {https://ui.adsabs.harvard.edu/abs/2021PhRvD.104b3007D}
}

@ARTICLE{Cai2021,
       author = {{Cai}, C. and {Xiong}, S.~L. and {Li}, C.~K. and {Liu}, C.~Z. and {Zhang}, S.~N. and {Li}, X.~B. and {Song}, L.~M. and {Li}, B. and {Xiao}, S. and {Yi}, Q.~B. and {Zhu}, Y. and {Zheng}, Y.~G. and {Chen}, W. and {Luo}, Q. and {Huang}, Y. and {Song}, X.~Y. and {Zhao}, H.~S. and {Zhao}, Y. and {Zhang}, Z. and {Bu}, Q.~C. and {Cao}, X.~L. and {Chang}, Z. and {Chen}, L. and {Chen}, T.~X. and {Chen}, Y.~B. and {Chen}, Y. and {Chen}, Y.~P. and {Cui}, W.~W. and {Du}, Y.~Y. and {Gao}, G.~H. and {Gao}, H. and {Ge}, M.~Y. and {Gu}, Y.~D. and {Guan}, J. and {Guo}, C.~C. and {Han}, D.~W. and {Huo}, J. and {Jia}, S.~M. and {Jiang}, W.~C. and {Jin}, J. and {Kong}, L.~D. and {Li}, G. and {Li}, T.~P. and {Li}, W. and {Li}, X. and {Li}, X.~F. and {Li}, Z.~W. and {Liang}, X.~H. and {Liao}, J.~Y. and {Liu}, B.~S. and {Liu}, H.~W. and {Liu}, H.~X. and {Liu}, X.~J. and {Lu}, F.~J. and {Lu}, X.~F. and {Luo}, T. and {Ma}, R.~C. and {Ma}, X. and {Meng}, B. and {Nang}, Y. and {Nie}, J.~Y. and {Ou}, G. and {Qu}, J.~L. and {Ren}, X.~Q. and {Sai}, N. and {Sun}, L. and {Tan}, Y. and {Tao}, L. and {Tuo}, Y.~L. and {Wang}, C. and {Wang}, L.~J. and {Wang}, P.~J. and {Wang}, W.~S. and {Wang}, Y.~S. and {Wen}, X.~Y. and {Wu}, B.~B. and {Wu}, B.~Y. and {Wu}, M. and {Xiao}, G.~C. and {Xu}, Y.~P. and {Yang}, R.~J. and {Yang}, S. and {Yang}, Y.~J. and {Yang}, Y.~R. and {Yang}, X.~J. and {Yin}, Q.~Q. and {You}, Y. and {Zhang}, F. and {Zhang}, H.~M. and {Zhang}, J. and {Zhang}, P. and {Zhang}, S. and {Zhang}, W.~C. and {Zhang}, W. and {Zhang}, Y.~F. and {Zhang}, Y.~H. and {Zhao}, X.~F. and {Zheng}, S.~J. and {Zhou}, D.~K.},
        title = "{Search for gamma-ray bursts and gravitational wave electromagnetic counterparts with High Energy X-ray Telescope of Insight-HXMT}",
      journal = {\mnras},
         year = 2021,
        month = dec,
       volume = {508},
       number = {3},
        pages = {3910-3920},
          doi = {10.1093/mnras/stab2760},
archivePrefix = {arXiv},
       eprint = {2109.12270},
 primaryClass = {astro-ph.HE},
       adsurl = {https://ui.adsabs.harvard.edu/abs/2021MNRAS.508.3910C}
}

@ARTICLE{Meegan2009,
       author = {{Meegan}, Charles and {Lichti}, Giselher and {Bhat}, P.~N. and {Bissaldi}, Elisabetta and {Briggs}, Michael S. and {Connaughton}, Valerie and {Diehl}, Roland and {Fishman}, Gerald and {Greiner}, Jochen and {Hoover}, Andrew S. and {van der Horst}, Alexander J. and {von Kienlin}, Andreas and {Kippen}, R. Marc and {Kouveliotou}, Chryssa and {McBreen}, Sheila and {Paciesas}, W.~S. and {Preece}, Robert and {Steinle}, Helmut and {Wallace}, Mark S. and {Wilson}, Robert B. and {Wilson-Hodge}, Colleen},
        title = "{The Fermi Gamma-ray Burst Monitor}",
      journal = {\apj},
         year = 2009,
        month = sep,
       volume = {702},
       number = {1},
        pages = {791-804},
          doi = {10.1088/0004-637X/702/1/791},
archivePrefix = {arXiv},
       eprint = {0908.0450},
 primaryClass = {astro-ph.IM},
       adsurl = {https://ui.adsabs.harvard.edu/abs/2009ApJ...702..791M}
}

@ARTICLE{Lin2020a,
       author = {{Lin}, Lin and {G{\"o}{\u{g}}{\"u}{\c{s}}}, Ersin and {Roberts}, Oliver J. and {Kouveliotou}, Chryssa and {Kaneko}, Yuki and {van der Horst}, Alexander J. and {Younes}, George},
        title = "{Burst Properties of the Most Recurring Transient Magnetar SGR J1935+2154}",
      journal = {\apj},
         year = 2020,
        month = apr,
       volume = {893},
       number = {2},
          eid = {156},
        pages = {156},
          doi = {10.3847/1538-4357/ab818f},
archivePrefix = {arXiv},
       eprint = {2003.10582},
 primaryClass = {astro-ph.HE},
       adsurl = {https://ui.adsabs.harvard.edu/abs/2020ApJ...893..156L}
}

@ARTICLE{Lin2020b,
       author = {{Lin}, Lin and {G{\"o}{\u{g}}{\"u}{\c{s}}}, Ersin and {Roberts}, Oliver J. and {Baring}, Matthew G. and {Kouveliotou}, Chryssa and {Kaneko}, Yuki and {van der Horst}, Alexander J. and {Younes}, George},
        title = "{Fermi/GBM View of the 2019 and 2020 Burst Active Episodes of SGR J1935+2154}",
      journal = {\apjl},
         year = 2020,
        month = oct,
       volume = {902},
       number = {2},
          eid = {L43},
        pages = {L43},
          doi = {10.3847/2041-8213/abbefe},
archivePrefix = {arXiv},
       eprint = {2010.02767},
 primaryClass = {astro-ph.HE},
       adsurl = {https://ui.adsabs.harvard.edu/abs/2020ApJ...902L..43L}
}

@ARTICLE{Lin2020c,
       author = {{Lin}, L. and {Zhang}, C.~F. and {Wang}, P. and {Gao}, H. and {Guan}, X. and {Han}, J.~L. and {Jiang}, J.~C. and {Jiang}, P. and {Lee}, K.~J. and {Li}, D. and {Men}, Y.~P. and {Miao}, C.~C. and {Niu}, C.~H. and {Niu}, J.~R. and {Sun}, C. and {Wang}, B.~J. and {Wang}, Z.~L. and {Xu}, H. and {Xu}, J.~L. and {Xu}, J.~W. and {Yang}, Y.~H. and {Yang}, Y.~P. and {Yu}, W. and {Zhang}, B. and {Zhang}, B. -B. and {Zhou}, D.~J. and {Zhu}, W.~W. and {Castro-Tirado}, A.~J. and {Dai}, Z.~G. and {Ge}, M.~Y. and {Hu}, Y.~D. and {Li}, C.~K. and {Li}, Y. and {Li}, Z. and {Liang}, E.~W. and {Jia}, S.~M. and {Querel}, R. and {Shao}, L. and {Wang}, F.~Y. and {Wang}, X.~G. and {Wu}, X.~F. and {Xiong}, S.~L. and {Xu}, R.~X. and {Yang}, Y. -S. and {Zhang}, G.~Q. and {Zhang}, S.~N. and {Zheng}, T.~C. and {Zou}, J. -H.},
        title = "{No pulsed radio emission during a bursting phase of a Galactic magnetar}",
      journal = {\nat},
         year = 2020,
        month = nov,
       volume = {587},
       number = {7832},
        pages = {63-65},
          doi = {10.1038/s41586-020-2839-y},
archivePrefix = {arXiv},
       eprint = {2005.11479},
 primaryClass = {astro-ph.HE},
       adsurl = {https://ui.adsabs.harvard.edu/abs/2020Natur.587...63L}
}

@ARTICLE{Mereghetti2020,
       author = {{Mereghetti}, S. and {Savchenko}, V. and {Ferrigno}, C. and {G{\"o}tz}, D. and {Rigoselli}, M. and {Tiengo}, A. and {Bazzano}, A. and {Bozzo}, E. and {Coleiro}, A. and {Courvoisier}, T.~J. -L. and {Doyle}, M. and {Goldwurm}, A. and {Hanlon}, L. and {Jourdain}, E. and {von Kienlin}, A. and {Lutovinov}, A. and {Martin-Carrillo}, A. and {Molkov}, S. and {Natalucci}, L. and {Onori}, F. and {Panessa}, F. and {Rodi}, J. and {Rodriguez}, J. and {S{\'a}nchez-Fern{\'a}ndez}, C. and {Sunyaev}, R. and {Ubertini}, P.},
        title = "{INTEGRAL Discovery of a Burst with Associated Radio Emission from the Magnetar SGR 1935+2154}",
      journal = {\apjl},
         year = 2020,
        month = aug,
       volume = {898},
       number = {2},
          eid = {L29},
        pages = {L29},
          doi = {10.3847/2041-8213/aba2cf},
archivePrefix = {arXiv},
       eprint = {2005.06335},
 primaryClass = {astro-ph.HE},
       adsurl = {https://ui.adsabs.harvard.edu/abs/2020ApJ...898L..29M}
}

@ARTICLE{Younes2020,
       author = {{Younes}, George and {G{\"u}ver}, Tolga and {Kouveliotou}, Chryssa and {Baring}, Matthew G. and {Hu}, Chin-Ping and {Wadiasingh}, Zorawar and {Begi{\c{c}}arslan}, Beste and {Enoto}, Teruaki and {G{\"o}{\u{g}}{\"u}{\c{s}}}, Ersin and {Lin}, Lin and {Harding}, Alice K. and {van der Horst}, Alexander J. and {Majid}, Walid A. and {Guillot}, Sebastien and {Malacaria}, Christian},
        title = "{NICER View of the 2020 Burst Storm and Persistent Emission of SGR 1935+2154}",
      journal = {\apjl},
         year = 2020,
        month = dec,
       volume = {904},
       number = {2},
          eid = {L21},
        pages = {L21},
          doi = {10.3847/2041-8213/abc94c},
archivePrefix = {arXiv},
       eprint = {2009.07886},
 primaryClass = {astro-ph.HE},
       adsurl = {https://ui.adsabs.harvard.edu/abs/2020ApJ...904L..21Y}
}

@ARTICLE{Blackburn2015,
       author = {{Blackburn}, L. and {Briggs}, M.~S. and {Camp}, J. and {Christensen}, N. and {Connaughton}, V. and {Jenke}, P. and {Remillard}, R.~A. and {Veitch}, J.},
        title = "{High-Energy Electromagnetic Offline Follow-Up of Ligo-Virgo Gravitational-Wave Binary Coalescence Candidate Events}",
      journal = {\apjs},
         year = 2015,
        month = mar,
       volume = {217},
       number = {1},
          eid = {8},
        pages = {8},
          doi = {10.1088/0067-0049/217/1/8},
archivePrefix = {arXiv},
       eprint = {1410.0929},
 primaryClass = {astro-ph.HE},
       adsurl = {https://ui.adsabs.harvard.edu/abs/2015ApJS..217....8B}
}

@ARTICLE{Briggs2013,
       author = {{Briggs}, Michael S. and {Xiong}, Shaolin and {Connaughton}, Valerie and {Tierney}, Dave and {Fitzpatrick}, Gerard and {Foley}, Suzanne and {Grove}, J. Eric and {Chekhtman}, Alexandre and {Gibby}, Melissa and {Fishman}, Gerald J. and {McBreen}, Shelia and {Chaplin}, Vandiver L. and {Guiriec}, Sylvain and {Layden}, Emily and {Bhat}, P.~N. and {Hughes}, Maximilian and {Greiner}, Jochen and {Kienlin}, Andreas and {Kippen}, R. Marc and {Meegan}, Charles A. and {Paciesas}, William S. and {Preece}, Robert D. and {Wilson-Hodge}, Colleen and {Holzworth}, Robert H. and {Hutchins}, Michael L.},
        title = "{Terrestrial gamma-ray flashes in the Fermi era: Improved observations and analysis methods}",
      journal = {Journal of Geophysical Research (Space Physics)},
         year = 2013,
        month = jun,
       volume = {118},
       number = {6},
        pages = {3805-3830},
          doi = {10.1002/jgra.50205},
       adsurl = {https://ui.adsabs.harvard.edu/abs/2013JGRA..118.3805B}
}

@ARTICLE{Goldstein2019,
       author = {{Goldstein}, Adam and {Hamburg}, Rachel and {Wood}, Joshua and {Hui}, C. Michelle and {Cleveland}, William H. and {Kocevski}, Daniel and {Littenberg}, Tyson and {Burns}, Eric and {Dal Canton}, Tito and {Veres}, Peter and {Mailyan}, Bagrat and {Malacaria}, Christian and {Briggs}, Michael S. and {Wilson-Hodge}, Colleen A.},
        title = "{Updates to the Fermi GBM Targeted Sub-threshold Search in Preparation for the Third Observing Run of LIGO/Virgo}",
      journal = {arXiv e-prints},
         year = 2019,
        month = mar,
          eid = {arXiv:1903.12597},
        pages = {arXiv:1903.12597},
archivePrefix = {arXiv},
       eprint = {1903.12597},
 primaryClass = {astro-ph.HE},
       adsurl = {https://ui.adsabs.harvard.edu/abs/2019arXiv190312597G}
}

@ARTICLE{Kaastra2017,
       author = {{Kaastra}, J.~S.},
        title = "{On the use of C-stat in testing models for X-ray spectra}",
      journal = {\aap},
         year = 2017,
        month = sep,
       volume = {605},
          eid = {A51},
        pages = {A51},
          doi = {10.1051/0004-6361/201629319},
archivePrefix = {arXiv},
       eprint = {1707.09202},
 primaryClass = {astro-ph.HE},
       adsurl = {https://ui.adsabs.harvard.edu/abs/2017A&A...605A..51K}
}

@ARTICLE{Lomb1976,
       author = {{Lomb}, N.~R.},
        title = "{Least-Squares Frequency Analysis of Unequally Spaced Data}",
      journal = {\apss},
         year = 1976,
        month = feb,
       volume = {39},
       number = {2},
        pages = {447-462},
          doi = {10.1007/BF00648343},
       adsurl = {https://ui.adsabs.harvard.edu/abs/1976Ap&SS..39..447L}
}

@ARTICLE{Scargle1982,
       author = {{Scargle}, J.~D.},
        title = "{Studies in astronomical time series analysis. II. Statistical aspects of spectral analysis of unevenly spaced data.}",
      journal = {\apj},
         year = 1982,
        month = dec,
       volume = {263},
        pages = {835-853},
          doi = {10.1086/160554},
       adsurl = {https://ui.adsabs.harvard.edu/abs/1982ApJ...263..835S}
}

@ARTICLE{VanderPlas2018,
       author = {{VanderPlas}, Jacob T.},
        title = "{Understanding the Lomb-Scargle Periodogram}",
      journal = {\apjs},
         year = 2018,
        month = may,
       volume = {236},
       number = {1},
          eid = {16},
        pages = {16},
          doi = {10.3847/1538-4365/aab766},
archivePrefix = {arXiv},
       eprint = {1703.09824},
 primaryClass = {astro-ph.IM},
       adsurl = {https://ui.adsabs.harvard.edu/abs/2018ApJS..236...16V}
}

@ARTICLE{IokaZhang2020,
       author = {{Ioka}, Kunihito and {Zhang}, Bing},
        title = "{A Binary Comb Model for Periodic Fast Radio Bursts}",
      journal = {\apjl},
         year = 2020,
        month = apr,
       volume = {893},
       number = {1},
          eid = {L26},
        pages = {L26},
          doi = {10.3847/2041-8213/ab83fb},
archivePrefix = {arXiv},
       eprint = {2002.08297},
 primaryClass = {astro-ph.HE},
       adsurl = {https://ui.adsabs.harvard.edu/abs/2020ApJ...893L..26I}
}

@ARTICLE{Lyutikov2020,
       author = {{Lyutikov}, Maxim and {Barkov}, Maxim V. and {Giannios}, Dimitrios},
        title = "{FRB Periodicity: Mild Pulsars in Tight O/B-star Binaries}",
      journal = {\apjl},
         year = 2020,
        month = apr,
       volume = {893},
       number = {2},
          eid = {L39},
        pages = {L39},
          doi = {10.3847/2041-8213/ab87a4},
archivePrefix = {arXiv},
       eprint = {2002.01920},
 primaryClass = {astro-ph.HE},
       adsurl = {https://ui.adsabs.harvard.edu/abs/2020ApJ...893L..39L}
}

@ARTICLE{Levin2020,
       author = {{Levin}, Yuri and {Beloborodov}, Andrei M. and {Bransgrove}, Ashley},
        title = "{Precessing Flaring Magnetar as a Source of Repeating FRB 180916.J0158+65}",
      journal = {\apjl},
         year = 2020,
        month = jun,
       volume = {895},
       number = {2},
          eid = {L30},
        pages = {L30},
          doi = {10.3847/2041-8213/ab8c4c},
archivePrefix = {arXiv},
       eprint = {2002.04595},
 primaryClass = {astro-ph.HE},
       adsurl = {https://ui.adsabs.harvard.edu/abs/2020ApJ...895L..30L}
}

@ARTICLE{ZanazziLai2020,
       author = {{Zanazzi}, J.~J. and {Lai}, Dong},
        title = "{Periodic Fast Radio Bursts with Neutron Star Free Precession}",
      journal = {\apjl},
         year = 2020,
        month = mar,
       volume = {892},
       number = {1},
          eid = {L15},
        pages = {L15},
          doi = {10.3847/2041-8213/ab7cdd},
archivePrefix = {arXiv},
       eprint = {2002.05752},
 primaryClass = {astro-ph.HE},
       adsurl = {https://ui.adsabs.harvard.edu/abs/2020ApJ...892L..15Z}
}

@ARTICLE{Tong2020,
       author = {{Tong}, Hao and {Wang}, Wei and {Wang}, Hong-Guang},
        title = "{Periodicity in fast radio bursts due to forced precession by a fallback disk}",
      journal = {Research in Astronomy and Astrophysics},
         year = 2020,
        month = sep,
       volume = {20},
       number = {9},
          eid = {142},
        pages = {142},
          doi = {10.1088/1674-4527/20/9/142},
archivePrefix = {arXiv},
       eprint = {2002.10265},
 primaryClass = {astro-ph.HE},
       adsurl = {https://ui.adsabs.harvard.edu/abs/2020RAA....20..142T}
}

@ARTICLE{YangZou2020,
       author = {{Yang}, Huan and {Zou}, Yuan-Chuan},
        title = "{Orbit-induced Spin Precession as a Possible Origin for Periodicity in Periodically Repeating Fast Radio Bursts}",
      journal = {\apjl},
         year = 2020,
        month = apr,
       volume = {893},
       number = {2},
          eid = {L31},
        pages = {L31},
          doi = {10.3847/2041-8213/ab800f},
archivePrefix = {arXiv},
       eprint = {2002.02553},
 primaryClass = {astro-ph.HE},
       adsurl = {https://ui.adsabs.harvard.edu/abs/2020ApJ...893L..31Y}
}

@ARTICLE{Stamatikos2014,
       author = {{Stamatikos}, M. and {Malesani}, D. and {Page}, K.~L. and {Sakamoto}, T.},
        title = "{GRB 140705A: Swift detection of a short burst.}",
      journal = {GRB Coordinates Network},
         year = 2014,
        month = jan,
       volume = {16520},
        pages = {1},
       adsurl = {https://ui.adsabs.harvard.edu/abs/2014GCN.16520....1S}
}

@ARTICLE{Younes2017,
       author = {{Younes}, George and {Kouveliotou}, Chryssa and {Jaodand}, Amruta and {Baring}, Matthew G. and {van der Horst}, Alexander J. and {Harding}, Alice K. and {Hessels}, Jason W.~T. and {Gehrels}, Neil and {Gill}, Ramandeep and {Huppenkothen}, Daniela and {Granot}, Jonathan and {G{\"o}{\u{g}}{\"u}{\c{s}}}, Ersin and {Lin}, Lin},
        title = "{X-Ray and Radio Observations of the Magnetar SGR J1935+2154 during Its 2014, 2015, and 2016 Outbursts}",
      journal = {\apj},
         year = 2017,
        month = oct,
       volume = {847},
       number = {2},
          eid = {85},
        pages = {85},
          doi = {10.3847/1538-4357/aa899a},
archivePrefix = {arXiv},
       eprint = {1702.04370},
 primaryClass = {astro-ph.HE},
       adsurl = {https://ui.adsabs.harvard.edu/abs/2017ApJ...847...85Y}
}

@ARTICLE{Li2021,
       author = {{Li}, C.~K. and {Lin}, L. and {Xiong}, S.~L. and {Ge}, M.~Y. and {Li}, X.~B. and {Li}, T.~P. and {Lu}, F.~J. and {Zhang}, S.~N. and {Tuo}, Y.~L. and {Nang}, Y. and {Zhang}, B. and {Xiao}, S. and {Chen}, Y. and {Song}, L.~M. and {Xu}, Y.~P. and {Liu}, C.~Z. and {Jia}, S.~M. and {Cao}, X.~L. and {Qu}, J.~L. and {Zhang}, S. and {Gu}, Y.~D. and {Liao}, J.~Y. and {Zhao}, X.~F. and {Tan}, Y. and {Nie}, J.~Y. and {Zhao}, H.~S. and {Zheng}, S.~J. and {Zheng}, Y.~G. and {Luo}, Q. and {Cai}, C. and {Li}, B. and {Xue}, W.~C. and {Bu}, Q.~C. and {Chang}, Z. and {Chen}, G. and {Chen}, L. and {Chen}, T.~X. and {Chen}, Y.~B. and {Chen}, Y.~P. and {Cui}, W. and {Cui}, W.~W. and {Deng}, J.~K. and {Dong}, Y.~W. and {Du}, Y.~Y. and {Fu}, M.~X. and {Gao}, G.~H. and {Gao}, H. and {Gao}, M. and {Gu}, Y.~D. and {Guan}, J. and {Guo}, C.~C. and {Han}, D.~W. and {Huang}, Y. and {Huo}, J. and {Jiang}, L.~H. and {Jiang}, W.~C. and {Jin}, J. and {Jin}, Y.~J. and {Kong}, L.~D. and {Li}, G. and {Li}, M.~S. and {Li}, W. and {Li}, X. and {Li}, X.~F. and {Li}, Y.~G. and {Li}, Z.~W. and {Liang}, X.~H. and {Liu}, B.~S. and {Liu}, G.~Q. and {Liu}, H.~W. and {Liu}, X.~J. and {Liu}, Y.~N. and {Lu}, B. and {Lu}, X.~F. and {Luo}, T. and {Ma}, X. and {Meng}, B. and {Ou}, G. and {Sai}, N. and {Shang}, R.~C. and {Song}, X.~Y. and {Sun}, L. and {Tao}, L. and {Wang}, C. and {Wang}, G.~F. and {Wang}, J. and {Wang}, W.~S. and {Wang}, Y.~S. and {Wen}, X.~Y. and {Wu}, B.~B. and {Wu}, B.~Y. and {Wu}, M. and {Xiao}, G.~C. and {Xu}, H. and {Yang}, J.~W. and {Yang}, S. and {Yang}, Y.~J. and {Yang}, Yi-Jung and {Yi}, Q.~B. and {Yin}, Q.~Q. and {You}, Y. and {Zhang}, A.~M. and {Zhang}, C.~M. and {Zhang}, F. and {Zhang}, H.~M. and {Zhang}, J. and {Zhang}, T. and {Zhang}, W. and {Zhang}, W.~C. and {Zhang}, W.~Z. and {Zhang}, Y. and {Zhang}, Yue and {Zhang}, Y.~F. and {Zhang}, Y.~J. and {Zhang}, Z. and {Zhang}, Zhi and {Zhang}, Z.~L. and {Zhou}, D.~K. and {Zhou}, J.~F. and {Zhu}, Y. and {Zhu}, Y.~X. and {Zhuang}, R.~L.},
        title = "{HXMT identification of a non-thermal X-ray burst from SGR J1935+2154 and with FRB 200428}",
      journal = {Nature Astronomy},
         year = 2021,
        month = apr,
       volume = {5},
        pages = {378-384},
          doi = {10.1038/s41550-021-01302-6},
archivePrefix = {arXiv},
       eprint = {2005.11071},
 primaryClass = {astro-ph.HE},
       adsurl = {https://ui.adsabs.harvard.edu/abs/2021NatAs...5..378L}
}

@ARTICLE{Tavani2020,
       author = {{Tavani}, M. and {Ursi}, A. and {Verrecchia}, F. and {Casentini}, C. and {Pittori}, C. and {Pilia}, M. and {Cardillo}, M. and {Piano}, G. and {Bulgarelli}, A. and {Fioretti}, V. and {Parmiggiani}, N. and {Lucarelli}, F. and {Donnarumma}, I. and {Vercellone}, S. and {Gianotti}, F. and {Trifoglio}, M. and {Giuliani}, A. and {Mereghetti}, S. and {Caraveo}, P. and {Perotti}, F. and {Chen}, A. and {Argan}, A. and {Costa}, E. and {Del Monte}, E. and {Evangelista}, Y. and {Feroci}, M. and {Lazzarotto}, F. and {Lapshov}, I. and {Pacciani}, L. and {Soffitta}, P. and {Vittorini}, V. and {Di Cocco}, G. and {Fuschino}, F. and {Galli}, M. and {Labanti}, C. and {Marisaldi}, M. and {Pellizzoni}, A. and {Trois}, A. and {Barbiellini}, G. and {Vallazza}, E. and {Longo}, F. and {Morselli}, A. and {Picozza}, P. and {Prest}, M. and {Lipari}, P. and {Zanello}, D. and {Cattaneo}, P.~W. and {Rappoldi}, A. and {Ferrari}, A. and {Paoletti}, F. and {Antonelli}, A. and {Giommi}, P. and {Salotti}, L. and {Valentini}, G. and {D'Amico}, F.},
        title = "{AGILE detection of a hard X-ray burst in temporal coincidence with a radio burst from SGR 1935+2154}",
      journal = {The Astronomer's Telegram},
         year = 2020,
        month = apr,
       volume = {13686},
        pages = {1},
       adsurl = {https://ui.adsabs.harvard.edu/abs/2020ATel13686....1T}
}

@ARTICLE{Ridnaia2021,
       author = {{Ridnaia}, A. and {Svinkin}, D. and {Frederiks}, D. and {Bykov}, A. and {Popov}, S. and {Aptekar}, R. and {Golenetskii}, S. and {Lysenko}, A. and {Tsvetkova}, A. and {Ulanov}, M. and {Cline}, T.~L.},
        title = "{A peculiar hard X-ray counterpart of a Galactic fast radio burst}",
      journal = {Nature Astronomy},
         year = 2021,
        month = apr,
       volume = {5},
        pages = {372-377},
          doi = {10.1038/s41550-020-01265-0},
archivePrefix = {arXiv},
       eprint = {2005.11178},
 primaryClass = {astro-ph.HE},
       adsurl = {https://ui.adsabs.harvard.edu/abs/2021NatAs...5..372R}
}

@ARTICLE{Bochenek2020,
       author = {{Bochenek}, C.~D. and {Ravi}, V. and {Belov}, K.~V. and {Hallinan}, G. and {Kocz}, J. and {Kulkarni}, S.~R. and {McKenna}, D.~L.},
        title = "{A fast radio burst associated with a Galactic magnetar}",
      journal = {\nat},
         year = 2020,
        month = nov,
       volume = {587},
       number = {7832},
        pages = {59-62},
          doi = {10.1038/s41586-020-2872-x},
archivePrefix = {arXiv},
       eprint = {2005.10828},
 primaryClass = {astro-ph.HE},
       adsurl = {https://ui.adsabs.harvard.edu/abs/2020Natur.587...59B}
}

@ARTICLE{Israel2016,
       author = {{Israel}, G.~L. and {Esposito}, P. and {Rea}, N. and {Coti Zelati}, F. and {Tiengo}, A. and {Campana}, S. and {Mereghetti}, S. and {Rodriguez Castillo}, G.~A. and {G{\"o}tz}, D. and {Burgay}, M. and {Possenti}, A. and {Zane}, S. and {Turolla}, R. and {Perna}, R. and {Cannizzaro}, G. and {Pons}, J.},
        title = "{The discovery, monitoring and environment of SGR J1935+2154}",
      journal = {\mnras},
         year = 2016,
        month = apr,
       volume = {457},
       number = {4},
        pages = {3448-3456},
          doi = {10.1093/mnras/stw008},
archivePrefix = {arXiv},
       eprint = {1601.00347},
 primaryClass = {astro-ph.HE},
       adsurl = {https://ui.adsabs.harvard.edu/abs/2016MNRAS.457.3448I}
}

@ARTICLE{Zhang2021,
       author = {{Zhang}, G.~Q. and {Tu}, Zuo-Lin and {Wang}, F.~Y.},
        title = "{Possible Periodic Activity in the Short Bursts of SGR 1806-20: Connection to Fast Radio Bursts}",
      journal = {\apj},
         year = 2021,
        month = mar,
       volume = {909},
       number = {1},
          eid = {83},
        pages = {83},
          doi = {10.3847/1538-4357/abdd27},
archivePrefix = {arXiv},
       eprint = {2101.07923},
 primaryClass = {astro-ph.HE},
       adsurl = {https://ui.adsabs.harvard.edu/abs/2021ApJ...909...83Z}
}

@ARTICLE{Petroff2016,
       author = {{Petroff}, E. and {Barr}, E.~D. and {Jameson}, A. and {Keane}, E.~F. and {Bailes}, M. and {Kramer}, M. and {Morello}, V. and {Tabbara}, D. and {van Straten}, W.},
        title = "{FRBCAT: The Fast Radio Burst Catalogue}",
      journal = {\pasa},
         year = 2016,
        month = sep,
       volume = {33},
          eid = {e045},
        pages = {e045},
          doi = {10.1017/pasa.2016.35},
archivePrefix = {arXiv},
       eprint = {1601.03547},
 primaryClass = {astro-ph.HE},
       adsurl = {https://ui.adsabs.harvard.edu/abs/2016PASA...33...45P}
}

@ARTICLE{Chatterjee2017,
       author = {{Chatterjee}, S. and {Law}, C.~J. and {Wharton}, R.~S. and {Burke-Spolaor}, S. and {Hessels}, J.~W.~T. and {Bower}, G.~C. and {Cordes}, J.~M. and {Tendulkar}, S.~P. and {Bassa}, C.~G. and {Demorest}, P. and {Butler}, B.~J. and {Seymour}, A. and {Scholz}, P. and {Abruzzo}, M.~W. and {Bogdanov}, S. and {Kaspi}, V.~M. and {Keimpema}, A. and {Lazio}, T.~J.~W. and {Marcote}, B. and {McLaughlin}, M.~A. and {Paragi}, Z. and {Ransom}, S.~M. and {Rupen}, M. and {Spitler}, L.~G. and {van Langevelde}, H.~J.},
        title = "{A direct localization of a fast radio burst and its host}",
      journal = {\nat},
         year = 2017,
        month = jan,
       volume = {541},
       number = {7635},
        pages = {58-61},
          doi = {10.1038/nature20797},
archivePrefix = {arXiv},
       eprint = {1701.01098},
 primaryClass = {astro-ph.HE},
       adsurl = {https://ui.adsabs.harvard.edu/abs/2017Natur.541...58C}
}

@ARTICLE{Bannister2019,
       author = {{Bannister}, K.~W. and {Deller}, A.~T. and {Phillips}, C. and {Macquart}, J. -P. and {Prochaska}, J.~X. and {Tejos}, N. and {Ryder}, S.~D. and {Sadler}, E.~M. and {Shannon}, R.~M. and {Simha}, S. and {Day}, C.~K. and {McQuinn}, M. and {North-Hickey}, F.~O. and {Bhandari}, S. and {Arcus}, W.~R. and {Bennert}, V.~N. and {Burchett}, J. and {Bouwhuis}, M. and {Dodson}, R. and {Ekers}, R.~D. and {Farah}, W. and {Flynn}, C. and {James}, C.~W. and {Kerr}, M. and {Lenc}, E. and {Mahony}, E.~K. and {O'Meara}, J. and {Os{\l}owski}, S. and {Qiu}, H. and {Treu}, T. and {U}, V. and {Bateman}, T.~J. and {Bock}, D.~C. -J. and {Bolton}, R.~J. and {Brown}, A. and {Bunton}, J.~D. and {Chippendale}, A.~P. and {Cooray}, F.~R. and {Cornwell}, T. and {Gupta}, N. and {Hayman}, D.~B. and {Kesteven}, M. and {Koribalski}, B.~S. and {MacLeod}, A. and {McClure-Griffiths}, N.~M. and {Neuhold}, S. and {Norris}, R.~P. and {Pilawa}, M.~A. and {Qiao}, R. -Y. and {Reynolds}, J. and {Roxby}, D.~N. and {Shimwell}, T.~W. and {Voronkov}, M.~A. and {Wilson}, C.~D.},
        title = "{A single fast radio burst localized to a massive galaxy at cosmological distance}",
      journal = {Science},
         year = 2019,
        month = aug,
       volume = {365},
       number = {6453},
        pages = {565-570},
          doi = {10.1126/science.aaw5903},
archivePrefix = {arXiv},
       eprint = {1906.11476},
 primaryClass = {astro-ph.HE},
       adsurl = {https://ui.adsabs.harvard.edu/abs/2019Sci...365..565B}
}

@ARTICLE{Ravi2019,
       author = {{Ravi}, V. and {Catha}, M. and {D'Addario}, L. and {Djorgovski}, S.~G. and {Hallinan}, G. and {Hobbs}, R. and {Kocz}, J. and {Kulkarni}, S.~R. and {Shi}, J. and {Vedantham}, H.~K. and {Weinreb}, S. and {Woody}, D.~P.},
        title = "{A fast radio burst localized to a massive galaxy}",
      journal = {\nat},
         year = 2019,
        month = aug,
       volume = {572},
       number = {7769},
        pages = {352-354},
          doi = {10.1038/s41586-019-1389-7},
archivePrefix = {arXiv},
       eprint = {1907.01542},
 primaryClass = {astro-ph.HE},
       adsurl = {https://ui.adsabs.harvard.edu/abs/2019Natur.572..352R}
}

@ARTICLE{Macquart2020,
       author = {{Macquart}, J. -P. and {Prochaska}, J.~X. and {McQuinn}, M. and {Bannister}, K.~W. and {Bhandari}, S. and {Day}, C.~K. and {Deller}, A.~T. and {Ekers}, R.~D. and {James}, C.~W. and {Marnoch}, L. and {Os{\l}owski}, S. and {Phillips}, C. and {Ryder}, S.~D. and {Scott}, D.~R. and {Shannon}, R.~M. and {Tejos}, N.},
        title = "{A census of baryons in the Universe from localized fast radio bursts}",
      journal = {\nat},
         year = 2020,
        month = may,
       volume = {581},
       number = {7809},
        pages = {391-395},
          doi = {10.1038/s41586-020-2300-2},
archivePrefix = {arXiv},
       eprint = {2005.13161},
 primaryClass = {astro-ph.CO},
       adsurl = {https://ui.adsabs.harvard.edu/abs/2020Natur.581..391M}
}

@ARTICLE{Marcote2020,
       author = {{Marcote}, B. and {Nimmo}, K. and {Hessels}, J.~W.~T. and {Tendulkar}, S.~P. and {Bassa}, C.~G. and {Paragi}, Z. and {Keimpema}, A. and {Bhardwaj}, M. and {Karuppusamy}, R. and {Kaspi}, V.~M. and {Law}, C.~J. and {Michilli}, D. and {Aggarwal}, K. and {Andersen}, B. and {Archibald}, A.~M. and {Bandura}, K. and {Bower}, G.~C. and {Boyle}, P.~J. and {Brar}, C. and {Burke-Spolaor}, S. and {Butler}, B.~J. and {Cassanelli}, T. and {Chawla}, P. and {Demorest}, P. and {Dobbs}, M. and {Fonseca}, E. and {Giri}, U. and {Good}, D.~C. and {Gourdji}, K. and {Josephy}, A. and {Kirichenko}, A. Yu. and {Kirsten}, F. and {Landecker}, T.~L. and {Lang}, D. and {Lazio}, T.~J.~W. and {Li}, D.~Z. and {Lin}, H. -H. and {Linford}, J.~D. and {Masui}, K. and {Mena-Parra}, J. and {Naidu}, A. and {Ng}, C. and {Patel}, C. and {Pen}, U. -L. and {Pleunis}, Z. and {Rafiei-Ravandi}, M. and {Rahman}, M. and {Renard}, A. and {Scholz}, P. and {Siegel}, S.~R. and {Smith}, K.~M. and {Stairs}, I.~H. and {Vanderlinde}, K. and {Zwaniga}, A.~V.},
        title = "{A repeating fast radio burst source localized to a nearby spiral galaxy}",
      journal = {\nat},
         year = 2020,
        month = jan,
       volume = {577},
       number = {7789},
        pages = {190-194},
          doi = {10.1038/s41586-019-1866-z},
archivePrefix = {arXiv},
       eprint = {2001.02222},
 primaryClass = {astro-ph.HE},
       adsurl = {https://ui.adsabs.harvard.edu/abs/2020Natur.577..190M}
}

@ARTICLE{Michilli2018,
       author = {{Michilli}, D. and {Seymour}, A. and {Hessels}, J.~W.~T. and {Spitler}, L.~G. and {Gajjar}, V. and {Archibald}, A.~M. and {Bower}, G.~C. and {Chatterjee}, S. and {Cordes}, J.~M. and {Gourdji}, K. and {Heald}, G.~H. and {Kaspi}, V.~M. and {Law}, C.~J. and {Sobey}, C. and {Adams}, E.~A.~K. and {Bassa}, C.~G. and {Bogdanov}, S. and {Brinkman}, C. and {Demorest}, P. and {Fernandez}, F. and {Hellbourg}, G. and {Lazio}, T.~J.~W. and {Lynch}, R.~S. and {Maddox}, N. and {Marcote}, B. and {McLaughlin}, M.~A. and {Paragi}, Z. and {Ransom}, S.~M. and {Scholz}, P. and {Siemion}, A.~P.~V. and {Tendulkar}, S.~P. and {van Rooy}, P. and {Wharton}, R.~S. and {Whitlow}, D.},
        title = "{An extreme magneto-ionic environment associated with the fast radio burst source FRB 121102}",
      journal = {\nat},
         year = 2018,
        month = jan,
       volume = {553},
       number = {7687},
        pages = {182-185},
          doi = {10.1038/nature25149},
archivePrefix = {arXiv},
       eprint = {1801.03965},
 primaryClass = {astro-ph.HE},
       adsurl = {https://ui.adsabs.harvard.edu/abs/2018Natur.553..182M}
}

@ARTICLE{Platts2019,
       author = {{Platts}, E. and {Weltman}, A. and {Walters}, A. and {Tendulkar}, S.~P. and {Gordin}, J.~E.~B. and {Kandhai}, S.},
        title = "{A living theory catalogue for fast radio bursts}",
      journal = {\physrep},
         year = 2019,
        month = aug,
       volume = {821},
        pages = {1-27},
          doi = {10.1016/j.physrep.2019.06.003},
archivePrefix = {arXiv},
       eprint = {1810.05836},
 primaryClass = {astro-ph.HE},
       adsurl = {https://ui.adsabs.harvard.edu/abs/2019PhR...821....1P}
}

@ARTICLE{Spitler2016,
       author = {{Spitler}, L.~G. and {Scholz}, P. and {Hessels}, J.~W.~T. and {Bogdanov}, S. and {Brazier}, A. and {Camilo}, F. and {Chatterjee}, S. and {Cordes}, J.~M. and {Crawford}, F. and {Deneva}, J. and {Ferdman}, R.~D. and {Freire}, P.~C.~C. and {Kaspi}, V.~M. and {Lazarus}, P. and {Lynch}, R. and {Madsen}, E.~C. and {McLaughlin}, M.~A. and {Patel}, C. and {Ransom}, S.~M. and {Seymour}, A. and {Stairs}, I.~H. and {Stappers}, B.~W. and {van Leeuwen}, J. and {Zhu}, W.~W.},
        title = "{A repeating fast radio burst}",
      journal = {\nat},
         year = 2016,
        month = mar,
       volume = {531},
       number = {7593},
        pages = {202-205},
          doi = {10.1038/nature17168},
archivePrefix = {arXiv},
       eprint = {1603.00581},
 primaryClass = {astro-ph.HE},
       adsurl = {https://ui.adsabs.harvard.edu/abs/2016Natur.531..202S}
}

@ARTICLE{Marcote2017,
       author = {{Marcote}, B. and {Paragi}, Z. and {Hessels}, J.~W.~T. and {Keimpema}, A. and {van Langevelde}, H.~J. and {Huang}, Y. and {Bassa}, C.~G. and {Bogdanov}, S. and {Bower}, G.~C. and {Burke-Spolaor}, S. and {Butler}, B.~J. and {Campbell}, R.~M. and {Chatterjee}, S. and {Cordes}, J.~M. and {Demorest}, P. and {Garrett}, M.~A. and {Ghosh}, T. and {Kaspi}, V.~M. and {Law}, C.~J. and {Lazio}, T.~J.~W. and {McLaughlin}, M.~A. and {Ransom}, S.~M. and {Salter}, C.~J. and {Scholz}, P. and {Seymour}, A. and {Siemion}, A. and {Spitler}, L.~G. and {Tendulkar}, S.~P. and {Wharton}, R.~S.},
        title = "{The Repeating Fast Radio Burst FRB 121102 as Seen on Milliarcsecond Angular Scales}",
      journal = {\apjl},
         year = 2017,
        month = jan,
       volume = {834},
       number = {2},
          eid = {L8},
        pages = {L8},
          doi = {10.3847/2041-8213/834/2/L8},
archivePrefix = {arXiv},
       eprint = {1701.01099},
 primaryClass = {astro-ph.HE},
       adsurl = {https://ui.adsabs.harvard.edu/abs/2017ApJ...834L...8M}
}

@ARTICLE{Connor2016,
       author = {{Connor}, Liam and {Sievers}, Jonathan and {Pen}, Ue-Li},
        title = "{Non-cosmological FRBs from young supernova remnant pulsars}",
      journal = {\mnras},
         year = 2016,
        month = may,
       volume = {458},
       number = {1},
        pages = {L19-L23},
          doi = {10.1093/mnrasl/slv124},
archivePrefix = {arXiv},
       eprint = {1505.05535},
 primaryClass = {astro-ph.HE},
       adsurl = {https://ui.adsabs.harvard.edu/abs/2016MNRAS.458L..19C}
}

@ARTICLE{Cordes2016,
       author = {{Cordes}, J.~M. and {Wasserman}, Ira},
        title = "{Supergiant pulses from extragalactic neutron stars}",
      journal = {\mnras},
         year = 2016,
        month = mar,
       volume = {457},
       number = {1},
        pages = {232-257},
          doi = {10.1093/mnras/stv2948},
archivePrefix = {arXiv},
       eprint = {1501.00753},
 primaryClass = {astro-ph.HE},
       adsurl = {https://ui.adsabs.harvard.edu/abs/2016MNRAS.457..232C}
}

@ARTICLE{Katz2016,
       author = {{Katz}, J.~I.},
        title = "{How Soft Gamma Repeaters Might Make Fast Radio Bursts}",
      journal = {\apj},
         year = 2016,
        month = aug,
       volume = {826},
       number = {2},
          eid = {226},
        pages = {226},
          doi = {10.3847/0004-637X/826/2/226},
archivePrefix = {arXiv},
       eprint = {1512.04503},
 primaryClass = {astro-ph.HE},
       adsurl = {https://ui.adsabs.harvard.edu/abs/2016ApJ...826..226K}
}

@ARTICLE{Lyutikov2017,
       author = {{Lyutikov}, Maxim},
        title = "{Fast Radio Bursts{\textquoteright} Emission Mechanism: Implication from Localization}",
      journal = {\apjl},
         year = 2017,
        month = mar,
       volume = {838},
       number = {1},
          eid = {L13},
        pages = {L13},
          doi = {10.3847/2041-8213/aa62fa},
archivePrefix = {arXiv},
       eprint = {1701.02003},
 primaryClass = {astro-ph.HE},
       adsurl = {https://ui.adsabs.harvard.edu/abs/2017ApJ...838L..13L}
}

@ARTICLE{Metzger2017,
       author = {{Metzger}, Brian D. and {Berger}, Edo and {Margalit}, Ben},
        title = "{Millisecond Magnetar Birth Connects FRB 121102 to Superluminous Supernovae and Long-duration Gamma-Ray Bursts}",
      journal = {\apj},
         year = 2017,
        month = may,
       volume = {841},
       number = {1},
          eid = {14},
        pages = {14},
          doi = {10.3847/1538-4357/aa633d},
archivePrefix = {arXiv},
       eprint = {1701.02370},
 primaryClass = {astro-ph.HE},
       adsurl = {https://ui.adsabs.harvard.edu/abs/2017ApJ...841...14M}
}

@INPROCEEDINGS{Popov2010,
       author = {{Popov}, Sergey B. and {Postnov}, K.~A.},
        title = "{Hyperflares of SGRs as an engine for millisecond extragalactic radio bursts}",
    booktitle = {Evolution of Cosmic Objects through their Physical Activity},
         year = 2010,
       editor = {{Harutyunian}, H.~A. and {Mickaelian}, A.~M. and {Terzian}, Y.},
        month = nov,
        pages = {129-132},
archivePrefix = {arXiv},
       eprint = {0710.2006},
 primaryClass = {astro-ph},
       adsurl = {https://ui.adsabs.harvard.edu/abs/2010vaoa.conf..129P}
}

@misc{GbmDataTools,
      author = {Adam Goldstein and William H. Cleveland and Daniel Kocevski},
      title = {Fermi GBM Data Tools: v1.1.1},
      year = 2022,
      url = {https://fermi.gsfc.nasa.gov/ssc/data/analysis/gbm}
}

%\begin{thebibliography}{99}
%\end{thebibliography}

%%%%%%%%%%%%%%%%%%%%%%%%%%%%%%%%%%%%%%%%%%%%%%%%%%

%%%%%%%%%%%%%%%%% APPENDICES %%%%%%%%%%%%%%%%%%%%%

% \appendix

% \section{Some extra material}

% If you want to present additional material which would interrupt the flow of the main paper,
% it can be placed in an Appendix which appears after the list of references.

%%%%%%%%%%%%%%%%%%%%%%%%%%%%%%%%%%%%%%%%%%%%%%%%%%

% Don't change these lines
\bsp	% typesetting comment
\label{lastpage}
\end{document}